\documentclass[11pt,a4paper]{article}
\usepackage[top=2.5cm,left=2cm,footskip=1cm,marginparwidth=0cm]{geometry}

\usepackage[T1]{fontenc}
\usepackage{graphicx}	% Including figure files
\usepackage{amsmath}	% Advanced maths commands
\usepackage{amssymb}	% Extra maths symbols
\usepackage{hyperref}
\usepackage{upgreek}
\usepackage{gensymb}
\usepackage{xfrac}
\usepackage{breqn}

\usepackage[round]{natbib}

\raggedright
\usepackage{changepage}

\usepackage[aboveskip=1pt,labelfont=bf,labelsep=period,singlelinecheck=off]{caption}

\usepackage{lastpage,fancyhdr,graphicx}
\usepackage{epstopdf}
\begin{document}
\vspace*{0.35in}

% Please keep new commands to a minimum, and use \newcommand not \def to avoid
% overwriting existing commands. Example:
%\newcommand{\pcm}{\,cm$^{-2}$}	% per cm-squared
\newcommand{\angstrom}{\text{\normalfont\AA}}
\newcommand{\ergs}{erg\,s^{-1}\,cm^{-2}\,\angstrom^{-1}}
\newcommand{\ergsarcsec}{erg\,s^{-1}\,cm^{-2}\,\angstrom^{-1}\,arcsec^{-2}}
\newcommand{\magsqm}{mag_{SQM}\,arcsec^{-2}}
\newcommand{\watts}{W\,m^{-2}\,sr^{-1}\,\upmu m^{-1}}
\newcommand{\ZP}{$\mathsf{ZP}$}
\newcommand{\SFD}{$\mathsf{SFD}$}
\newcommand{\NSB}{<NSB>}

%%%%%%%%%%%%%%%%%%%%%%%%%%%%%%%%%%%%%%%%%%%%%%%%%%

% title goes here:
\begin{center}
{\Large
\textbf{
Circalunar variations of the night sky brightness -- an FFT\\[0.1in]
perspective on the impact of light pollution
}}
\end{center}
\bigskip

\begin{flushleft}

% authors go here:
Johannes Puschnig,$^{1,*}$
Stefan Wallner,$^{2,3}$
Thomas Posch$^{2}$
\\
\bigskip
* e-mail: johannes.puschnig@uni-bonn.de\\
\bigskip
1 Universit\"at Bonn, Argelander-Institut f\"ur Astronomie, Auf dem H\"ugel 71, D-53121 Bonn, Germany\\
2 Universit\"at Wien, Institut f\"ur Astrophysik, T\"urkenschanzstra{\ss}e 17, A-1180 Wien, Austria\\
3 ICA, Slovak Academy of Sciences, Dubravska cesta 9, 84503 Bratislava, Slovak Republic\\
\bigskip
Accepted for publication in MNRAS on Dec 11, 2019.\\

\end{flushleft}

\section*{Abstract}
Circa-monthly activity conducted by moonlight is observed in many species on Earth. Given the vast amount of artificial light at night (ALAN) that pollutes large areas around the globe, the synchronization to the circalunar cycle is often strongly perturbed. Using two-year data from a network of 23 photometers (Sky Quality Meters; SQM) in Austria (latitude $\sim$48$^{\degree}$), we quantify how light pollution impacts the recognition of the circalunar periodicity. We do so via frequency analysis of nightly mean sky brightnesses using Fast Fourier Transforms. A very tight linear relation between the mean zenithal night sky brightness (NSB) given in $\mathsf{\magsqm}$ and the amplitude of the circalunar signal is found, indicating that for sites with a mean zenithal NSB brighter than 16.5 $\mathsf{\magsqm}$ the lunar rhythm practically vanishes. This finding implies that the circalunar rhythm is still detectable (within the broad bandpass of the SQM) at most places around the globe, but its amplitude against the light polluted sky is strongly reduced. We find that the circalunar contrast in zenith is reduced compared to ALAN-free sites by factors of $\mathsf{\sfrac{1}{9}}$ in the state capital of Linz ($\sim$200,000 inhabitants) and $\mathsf{\sfrac{1}{3}}$ in small towns, e.g. Freistadt and Mattighofen, with less than 10,000 inhabitants. Only two of our sites, both situated in national parks (Bodinggraben and Z\"oblboden), show natural circalunar amplitudes. At our urban sites we further detect a strong seasonal signal that is linked to the amplification of anthropogenic skyglow during the winter months due to climatological conditions.

%%%%%%%%%%%%%%%%% BODY OF PAPER %%%%%%%%%%%%%%%%%%

%%%%%%%%%%%%%%%%%%%%%%%%%%%%%%%%%%%%%%%%%%%%%%%%%%%%%%%%%%%%%%%%%%%%%%%%%%%%%%%%%%%
%%%%%%%%%%%%%%%%%%%%%%%%%%%%%%%%%%%%%%%%%%%%%%%%%%%%%%%%%%%%%%%%%%%%%%%%%%%%%%%%%%%
%%%%%%%%%%%%%%%%%%%%%%%%%%%%%%%%%%%%%%%%%%%%%%%%%%%%%%%%%%%%%%%%%%%%%%%%%%%%%%%%%%%
%%%%%%%%%%%%%%%%%%%%%%%%%%%%%%%%%%%%%%%%%%%%%%%%%%%%%%%%%%%%%%%%%%%%%%%%%%%%%%%%%%%
%%%%%%%%%%%%%%%%%%%%%%%%%%%%%%%%%%%%%%%%%%%%%%%%%%%%%%%%%%%%%%%%%%%%%%%%%%%%%%%%%%%
%%%%%%%%%%%%%%%%%%%%%%%%%%%%%%%%%%%%%%%%%%%%%%%%%%%%%%%%%%%%%%%%%%%%%%%%%%%%%%%%%%%
% Introduction, motivation, short survey of the relevant literature (TP, SW)
\section{Introduction}\label{sec:intro}

%%%%%%%%%%%%%%%%%%%%%%%%%%%%%%%%%%%%%%%%%%%%%%%%%%
\subsection{Impact of moonlight on animals, plants and humans}
The Moon's synodic period of 29.5 days is its orbital time around the Earth required to show
the exact same lunar phase, i.e. for example the time span between two consecutive full moons.
The corresponding circalunar oscillation of the Moon's illumination impacts
many types of life on Earth, in particular in the context of reproduction cycles. Scientific work
on this topic dates back to the early 20$\mathsf{^{th}}$ century (e.g. \cite{Fox1924}) and it was likely
already recognized by fishermen in the antiquity -- due to practical implications --
that the size of (edible) gonads of sea urchins varies over the lunar month \citep{Raible2017}.

Later studies revealed that also predator-prey interactions change with moon phase and
illumination \citep{Clarke1983, Shimose2013}, giving advantages to either side,
depending on the context. More recently,
\cite{Fallows2016} studied interactions between white sharks (\textit{Carcharodon carcharias})
and Cape fur seals (\textit{Arctocephalus pusillus pusillus}). They found that
the shark attack frequency and seal capture success was significantly higher
at sunrise during periods of low lunar illumination.
In other species, the lunar cycle may control activity \citep{Kolb1992}, foraging,
habitat use and communication. See \cite{Kronfeld-Schor2013} for a recent review on those topics.
And for the golden rabbitfish (\textit{Siganus guttatus}), \cite{Takemura2006} found a direct
influence of moonlight intensity on changes in melatonin production.

Some animals are driven by the Moon in their orientation \citep{Papi1963,Frisch1993}, in particular
\cite{Dacke2011} found that dung beetles (\textit{carabaeus lamarcki}) use the polarization pattern around the Moon
as a compass for maintaining their travel direction.

\cite{Buenning1968} studied how different types of plants react on moonlight. They revealed
that plants may undergo leaf movements such that the intensity of lunar illumination
is reduced and disturbing effects caused by moonlight are eliminated. It was further shown that
illumination by moonlight may even promote flowering, e.g. \cite{Ben-Attia2016}
found flowering patterns in the cactus \textit{Cereus peruvianus} with a period of $\sim$29.5 days
and a correlation between moon phase and number and proportion of flowers in bloom. 
In aquatic systems, \cite{Zantke2013} first established that the marine worm
\textit{Platynereis dumerilii} possesses an endogenous circalunar clock and
\cite{Last2016} was one of the first to search for impacts on aquatic ecosystems
and especially noticed a vertical migration of
zooplankton which takes place in winter when the Moon is above the horizon at
the Arctic, fjord or other sea areas.

Besides animals and plants, the impact of moonlight on humans is still under debate, see for example
\cite{Zimecki2006} for a review.
However, many authors find evidence that women of reproductive age do follow the circalunar
rhythm \citep{Reinberg2016}, especially the ovulation seems to accumulate around
new moon \citep{Law1986}.
%Nonetheless, influences on human behavior seem not to be validated and still have to be part of future studies.

What about the effect of lunar illumination on human sleep?
On one hand,
\cite{Cajochen2013} find that at full moon the electroencephalogram
delta activity during the deep sleep phase is 30 percent decreased
and that the sleep duration is reduced by 20 minutes, but
on the other hand, \cite{Cordi2014} find no such significant effects depending on lunar cycle.
However, \cite{Cajochen2014reply} pointed out that the volunteers tested by \cite{Cordi2014} were not
synchronised with respect to their own natural sleep timing, which may have led to low signal-to-noise
in their result.

Given the plethora of studies about how moonlight impacts various species on Earth, our work
on how artificial light at night (ALAN) impacts the recognizability of the lunar cycle, seems to be a
timely matter, as ALAN thus presumably perturbs those species in manifold ways (reproduction,
predator-prey interaction, activity, orientation, and so forth) as well.

%%%%%%%%%%%%%%%%%%%%%%%%%%%%%%%%%%%%%%%%%%%%%%%%%%
\subsection{Moonlight versus light pollution}
%As summarized in the previous section, the influence of moonlight on beings on our planet is well documented,
%including multiple disciplines such as biology, ecology, medicine and astronomy.
Given the fact that ALAN is ever increasing on a global scale \citep{Kyba2015}, effects of
artificial light on organisms and ecosystems have gained increasingly more attention in recent years
\citep{Hoelker2010,Gaston2015,Manfrin2017}. It has become clear that ALAN indeed has multifaceted
consequences for flora and fauna, see e.g. \cite{Schroer2016} for a review.

However, despite the fact that the influence of moonlight on beings on our planet is well documented (see previous section),
and that ALAN's impact on various organisms and ecosystems have been demonstrated by many authors, 
to date no studies were performed to investigate ALAN's impact on processes that rely on 
synchronization with moonlight. The reason for this knowledge gap is
that the community is lacking a quantification of the strength of the circalunar rhythm in dependence
of the level of light pollution \citep{Davies2013}. With this work, we aim to provide such a quantification,
i.e. a simple empirical relation between the mean night sky brightness and the amplitude of the
circalunar rhythm, allowing the knowledge gap to be filled in the near future.

\section{Locations and methods of our measurements} \label{sec:locmeth}
\begin{figure*}
        \begin{center}
        \includegraphics[width=\textwidth]{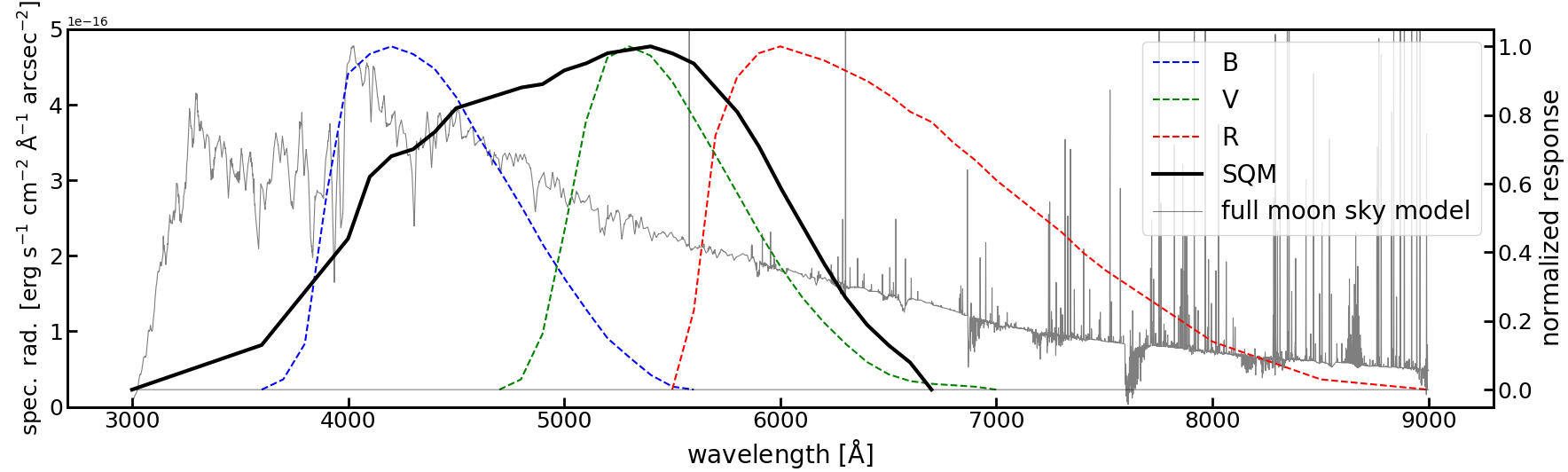}
        \caption[sky model spectrum and transmission curves]{Skycalc radiance model for the zenith with the
        full moon at 45 degree altitude. Transmission curves of the SQM and Bessel B, V, R filters are overplotted.}
        \label{fig:sqmtrans}
        \end{center}
\end{figure*}

The present paper is primarily based on zenithal night sky brightness (NSB)
measurements taken with Sky Quality Meters (model SQM-LE). These are
photodiode based devices with an optical element on the front that narrows down the
field of view to a Gaussian-like cone with a full-with-at-half-maximum
of $\sim$20$^\degree$. Its effective bandpass ranges from approximately
400--650nm (see Figure \ref{fig:sqmtrans}). Technical characterisation and testing was done by \cite{Cinzano2005}
and details about the absolute radiometric calibration are found in \cite{Bara2019}.

Our SQM measurements have been carried out in the Austrian county of Upper Austria at 23 sites,
distributed over the whole area of this county and ranging from its capital city
Linz -- which has very bright skies -- to very remote locations such as Krippenstein on the Dachstein
plateau ($\sim$2000m above sea level). See Table \ref{tab:sqmsites} for station codes and
geographic coordinates. Measurements are taken in an automated way, with SQM
devices located in weather-proof housings. The network of SQMs is run by the provincial
government of Upper Austria. Starting with six devices in 2014, it has grown to 23 SQMs by 2016.
A detailed description of the individual sites, their exact locations
and light pollution levels are found in \cite{Posch2018}.

Measurements are taken every minute, thus a huge amount of data is generated every night.
However, the present data analysis is based on the {\textit{mean} nocturnal NSBs} (\NSB) and is
restricted to data obtained during the years 2016--2017.
For each night and SQM site, we calculate \NSB\ as  arithmetic means of the
minute-by-minute SQM readings for each night. The data series is further constrained for solar
elevations below -15 degrees, in the same way as described by \cite{Posch2018}. The contribution
of scattered sunlight to the night sky brightness is negligible below this altitude.
Our \NSB\ measurements range from 17.3--21.0 $\mathsf{\magsqm}$.

We stress that
we do not apply any further filtering, i.e. data obtained when the sky was cloudy are included.

\begin{table*}
\centering
\caption[Basic information of the 23 measurement sites]{Basic information of the 23 SQM
locations in Upper Austria,
categorized into urban, intermediate and rural sites, as done in \cite{Posch2018}.}
\label{tab:sqmsites}
\begin{tabular}{lllll}
Code & Name                      & Latitude N   & Longitude E      & Elevation {[}m{]}   \\
     &                           &              &                  & (above sea level) \\
\hline\hline
     &                           &              & \textit{urban}                &                     \\
LSM  & Linz, Schlossmuseum        & N 48 18 19 & E 14 16 58     & 287                 \\
LGO  & Linz, G\"othestra{\ss}e    & N 48 18 19 & E 14 18 30     & 259                 \\
LSW  & Linz, Sternwarte           & N 48 17 36 & E 14 16 6      & 336                 \\
STY  & Steyr                     & N 48 2 57  & E 14 26 32     & 307                 \\
STW  & Steyregg-Weih             & N 48 17 19 & E 14 21 13     & 331                 \\
TRA  & Traun                     & N 48 14 8  & E 14 15 11     & 269                 \\
WEL  & Wels, Rathaus              & N 48 9 23  & E 14 1 29      & 317                 \\
     &                           &              &  \textit{intermediate} &                     \\
BRA  & Braunau                   & N 48 15 40 & E 13 2 41      & 351                 \\
GRI  & Grieskirchen              & N 48 14 4  & E 13 49 33     & 336                 \\
FRE  & Freistadt                 & N 48 30 33 & E 14 30 7      & 512                 \\
MAT  & Mattighofen               & N 48 5 50  & E 13 9 6       & 454                 \\
PAS  & Pasching                  & N 48 15 31 & E 14 12 36     & 292                 \\
VOE  & V\"ocklabruck               & N 48 0 21  & E 13 38 43     & 434                 \\
     &                           &              &  \textit{rural}        &                     \\
BOD  & Nationalpark Bodinggraben & N 47 47 31 & E 14 23 38     & 641                 \\
FEU  & Feuerkogel                & N 47 48 57 & E 13 43 15     & 1628                \\
GIS  & Giselawarte               & N 48 23 3  & E 14 15 11     & 902                 \\
GRU  & Gr\"unbach                  & N 48 31 50 & E 14 34 30     & 918                 \\
KID  & Kirchschlag-Davidschlag   & N 48 26 31 & E 14 16 26     & 813                 \\
KRI  & Krippenstein              & N 47 31 23 & E 13 41 36     & 2067                \\
LOS  & Losenstein, Hohe Dirn      & N 47 54 22 & E 14 24 40     & 982                 \\
MUN  & M\"unzkirchen               & N 48 28 45 & E 13 33 29     & 486                 \\
ULI  & Ulrichsberg, Sch\"oneben     & N 48 42 20 & E 13 56 44     & 935                 \\
ZOE  & Nationalpark Z\"obloden    & N 47 50 18 & E 14 26 28     & 899                
\end{tabular}
\end{table*}

%%%%%%%%%%%%%%%%%%%%%%%%%%%%%%%%%%%%%%%%%%%%%%%%%%%%%%%%%%%%%%%%%%%%%%%%%%%%%%%%%%%
%%%%%%%%%%%%%%%%%%%%%%%%%%%%%%%%%%%%%%%%%%%%%%%%%%%%%%%%%%%%%%%%%%%%%%%%%%%%%%%%%%%
%%%%%%%%%%%%%%%%%%%%%%%%%%%%%%%%%%%%%%%%%%%%%%%%%%%%%%%%%%%%%%%%%%%%%%%%%%%%%%%%%%%
%%%%%%%%%%%%%%%%%%%%%%%%%%%%%%%%%%%%%%%%%%%%%%%%%%%%%%%%%%%%%%%%%%%%%%%%%%%%%%%%%%%
%%%%%%%%%%%%%%%%%%%%%%%%%%%%%%%%%%%%%%%%%%%%%%%%%%%%%%%%%%%%%%%%%%%%%%%%%%%%%%%%%%%
\section{Synthetic models of ground illumination by the Moon and total zenithal night sky brightness}
\label{sec:models}
In order to study ALAN's impact on the lunar rhythm, we first want
to understand the amplitude of the \textit{naturally} occuring oscillation of moonlight at our sites,
i.e. without any contribution of anthropogenic light at night.
We do so using two models. The first one, describing the lunar variation of \textit{ground illumination}
in units of \textit{lux} and the second one, describing the naturally occuring
variation of zenithal night sky brightness in units of $\mathsf{\magsqm}$, which may be approximately
converted to luminance in units of $\mathsf{cd\ m^{-2}}$ \citep{Bara2016,Bara2017},
using Equation \ref{eq:unihedron}. Despite the fact that this formula is widely used
to estimate luminances from SQM magnitudes, it was originally derived by \cite{Garstang1986} for
Johnson V-band magnitudes, and is thus just an approximation.

\begin{dmath}
   Luminance\ [cd/m^2] = 10.8 \times 10^4 \times 10^{(-0.4\ \times\ [mag\ arcsec^{-2}])}
   \label{eq:unihedron}
\end{dmath}

%%%%%%%%%%%%%%%%%%%%%%%%%%%%%%%%%%%%%%%%%%%%%%%%%%%%%%%%%%%%%%%%%%%%%%%%%%%%%%%%%%%
\subsection{Simplified model of ground illumination by the Moon}\label{sec:illuminancemodel}

\subsubsection{General solution for all possible values}
\begin{figure}
        \begin{center}
        \includegraphics[width=0.7\columnwidth]{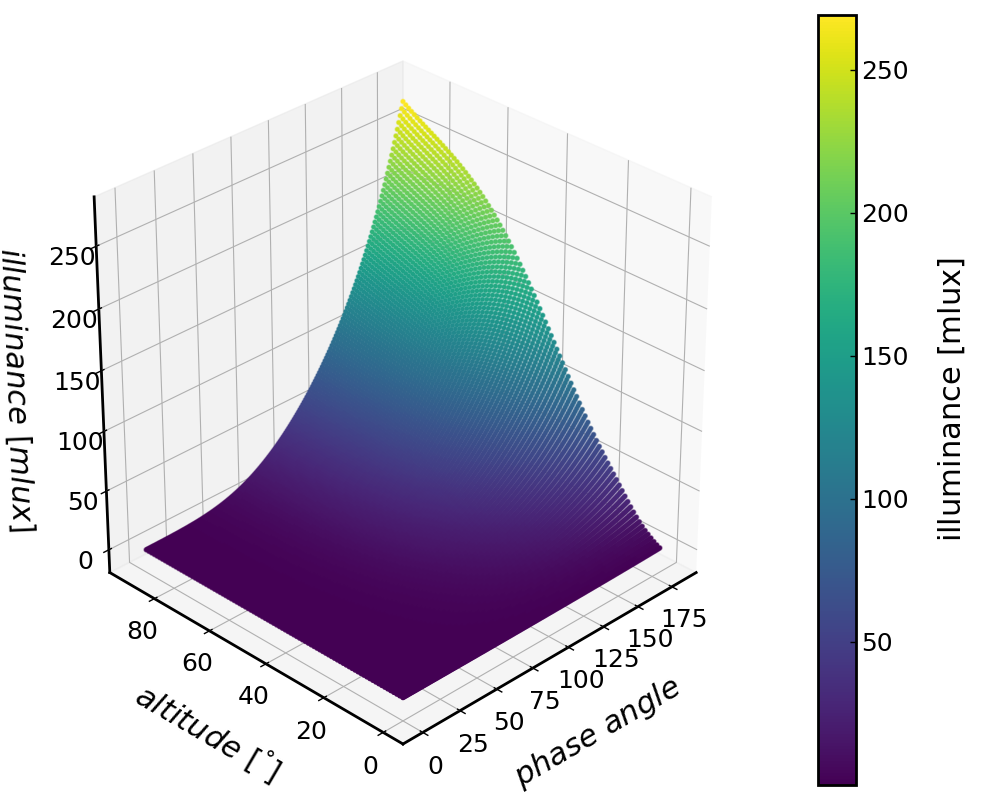}
        \caption[lunar illuminance model]{Results of the illuminance model as developed by \cite{Seidelmann1992}
        for all phase angles and altitudes of the Moon.}
        \label{fig:moonmodel}
        \end{center}
\end{figure}

To obtain insight into the contribution of moonlight to the total ground illumination, we
make use of the
moonlight model by \cite{Seidelmann1992}.
The model does not take into account contributions of the skyglow, stars or airglow
and thus depend only on two parameters: the mean altitude of the Moon (0-90$^{\circ}$, where 90$^{\circ}$ is the zenith)
and the Moon's phase angle (0-180$^{\circ}$, where 180$^{\circ}$ is full moon). Note that the parallax value
is neglected due to its very small contribution.
We show the whole parameter space covered by the model in Figure \ref{fig:moonmodel}.

Unsurprisingly, the amount of ground illuminance is highest when the full moon can be observed in the zenith. It is recognized that with increasing phase angle and altitude the illumination does not increase linear,
but rather exponential, owed to the fact that the transmittance of the atmosphere is proportional to
$e^{-\frac{\tau}{cos(z)}}$, with $\tau$ being the optical depth and $z$ the zenith distance. 
Some values which underline this can be found in Table \ref{tab:moonmodel}. In fact, for zenithal positions,
there is a factor $\sim$8.4 in ground illumination between full and half moon.
The large difference is caused by \textit{coherent backscattering} or \textit{opposition surge} \citep{Hapke1998}.
In this phenomenon portions of waves traveling along same paths but in opposite directions,
interfere constructively with each other, causing a peak at zero phase (full moon).
%%% originally added by SW
%When conducting, that brightness of the half moon decreases by a factor of 10, compared to full moon,
%the resulting value underlines this. Speaking of which, this effect can be explained by its low surface
%albedo of 0.12 and the decrease of reflected light with increasing illumination angle, latter being zero
%at full moon.
%%%%%%%%%%%%%%

In case of full moon the illuminance varies by a factor of $\sim$166 between zenithal and horizontal positions.
Please note that horizontal position, i.e. altitude of zero degrees, is valid for the center of the
lunar disk. Hence, a slight ground illumination is visible and not zero.
The resulting values can be considered as theoretical only, since
every other possible light emitting source is neglected, 
even stars or the Milky Way are not taken into account. However, the model shows how the lunar position
and its phase angle are associated with the consequential ground illumination.

\begin{table}
    \centering
    \caption{Resulting values of the theoretical Moon model for half and full moons. ALT is the lunar altitude, PA is the lunar phase angle and $\mathsf{G_{ILL}}$ is the ground illumination caused only by moonlight.}
    \begin{tabular}{ccc}
    ALT [$^{\circ}$] & PA [$^{\circ}$] & $\mathsf{G_{ILL}}$ [$\mathsf{mlux}$]\\
    \hline \hline
    0 & 90 & 0.192\\
    0 & 180 & 1.622\\
    45 & 90 & 19.474\\
    45 & 180 & 164.059\\
    90 & 90 & 31.949\\
    90 & 180 & 269.153\\
    \end{tabular}
    \label{tab:moonmodel}
\end{table}

%%%%%%%%%%%%%%%%%%%%%%%%%%%%%%%%%%%%%%%%%%%%%%%%%%%%%%%%%%%%%%%%%%%%%%%%%%%%%%%%%%%
\subsubsection{Application of the illuminance model to one of our sites}\label{sec:appliillumodel}
\begin{figure}
        \begin{center}
        \includegraphics[width=0.7\columnwidth]{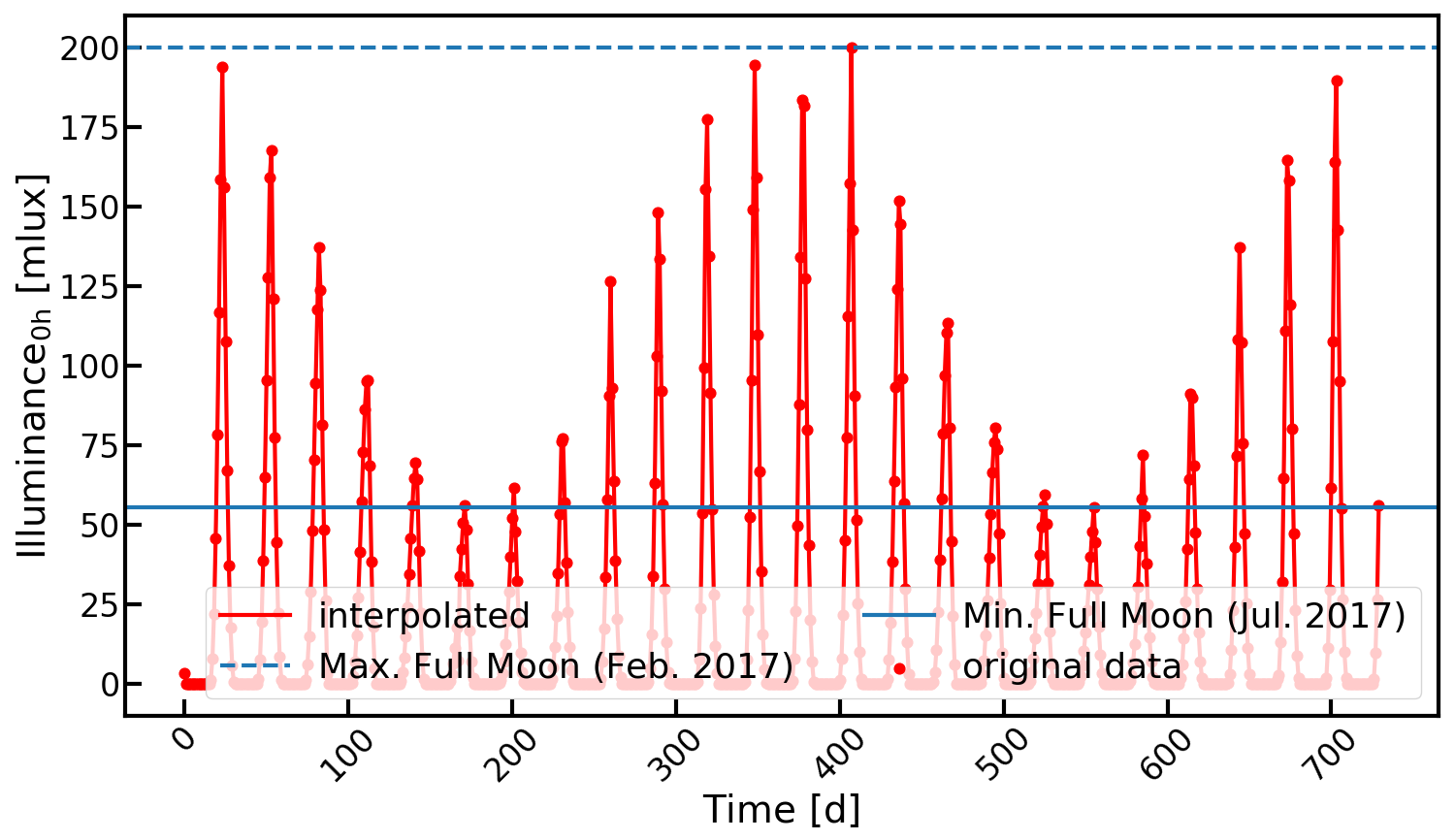}
        \caption{Ground illuminance (the Moon's contribution to it) evaluated for the location of V\"ocklabruck, Upper Austria, over
        the years 2016--2017. The high-frequency lunar cycle is modulated with a seasonal variation caused
        by the seasonal change of the Moon's altitude.}
        \label{fig:LUXVOE}
        \end{center}
\end{figure}

We now apply the model to one of our sites, namely VOE. To do so,
we first calculate
for the years 2016--2017
the altitude and phase of
the Moon \textit{at midnight} and then derive the corresponding illuminance through the model.
The result is shown in Figure \ref{fig:LUXVOE}.
A strong seasonal variation is recognized in the model, mainly caused by the changing altitude of
the ecliptic (and thus the Moon) between summer and winter, leading to a natural variation of the moon
illuminance by a factor of $\sim$3.6 (peak-to-peak for full moon).

%%%%%%%%%%%%%%%%%%%%%%%%%%%%%%%%%%%%%%%%%%%%%%%%%%%%%%%%%%%%%%%%%%%%%%%%%%%%%%%%%%%
\subsection{Zenithal night sky model for the SQM band}\label{sec:skymodel}

\subsubsection{The Cerro Paranal Advanced Sky Model}
Using the \textit{Cerro Paranal Advanced Sky Model} (Skycalc), we are able to compare
our zenithal SQM measurements to a synthetic sky model that is
cloud-free and takes into account several sources of light such as
scattered moonlight, starlight, molecular emission of the lower atmosphere and the airglow (upper atmosphere).

Skycalc was published by \cite{Noll2012} and \cite{Jones2013}, as part of an Austrian in-kind
contribution to the European Southern Observatory (ESO), e.g. ESO's exposure time calculator is based on it.
Skycalc's current version (2.0.4) also comes with a Python-based command line
interface\footnote{\url{https://www.eso.org/observing/etc/doc/skycalc/helpskycalccli.html}}.
However, the model used for our study is based on Skycalc 1.4.4, which was available through a
web interface only.
The main input parameters are zenith distance (or airmass) of the observation,
precipitable water vapor (PWV) and
monthly averaged solar flux. For the moon radiance component, the separation of Sun and Moon
as seen from Earth, the Moon-target separation, moon altitude over horizon and the Moon-Earth
distance are needed.

We have decided to make some simplifications, allowing us to
evaluate the model on a 2-dimensional parameter grid with vectors of (Sun-Moon-separation, moon
altitude) only. This is reasonable in our case, because the measurement devices we are using, the Sky
Quality Meters of type SQM-LE, are equipped with a front lens that narrows down the field of view
to a roughly 20 degree wide cone, pointed towards zenith. Hence, we only need to consider zenithal
night sky brightness. The two input parameters Moon-target separation and moon altitude can thus
be simplified to one parameter, with the former one being the moon zenith distance. We have further
decided to evaluate the model for a fixed PWV value of 5mm, a monthly averaged
solar flux of 130sfu and for a fixed mean Moon-Earth distance. These simplifications have practically
no influence on our results, since ALAN's contribution to our SQM measurements is magnitudes larger
than the natural variation caused by phenomena such as PWV or solar flux. However, variations
due to moon phase and height are fully treated by our
gridded model evaluation for the zenith.
Since the natural, cloudless sky brightness changes smoothly, a grid spacing of one degree in both
parameters (Sun-Moon-separation, moon altitude) was found to be sufficient.

The result is a synthetic (cloud-free) night sky spectrum for the target location
in units of $\mathsf{photons\,s^{-1}\,m^{-2}\,\upmu m^{-1}\,arcsec^{-2}}$, i.e. spectral radiance.
We first convert to $\mathsf{\ergsarcsec}$\ and then multiply with the SQM transmission curve as
published in \cite{Cinzano2005} and shown in Figure \ref{fig:sqmtrans}, together with transmission
curves of Bessel BVR filters.
The radiance within the SQM band is then calculated via integration over the wavelength axis.
Using a zeropoint (ZP) of -12.92 (Puschnig et al. in prep),
we finally convert to $\mathsf{\magsqm}$ via equation \ref{eq:radiance2mag}.

\begin{dmath}
    NSB\ [\magsqm] = -2.5 \times log_{10}(radiance\ [erg\ s^{-1}\ cm^{-2}\ arcsec^{-2}]) + ZP
    \label{eq:radiance2mag}
\end{dmath}

Note that Equation \ref{eq:radiance2mag} results to a modeled zenithal sky brightness of 21.87$\mathsf{\magsqm}$
for new moon, which is in agreement with SQM observations of remote, rural sites \citep{Posch2018,Bertolo2019,Bara2019}.
We further stress that the exact absolute value is not critical for the scientific results of the paper.

\subsubsection{Application of the Cerro Paranal Advanced Sky Model to one of our sites}\label{sec:appliskymodel}
\begin{figure}
        \begin{center}
        \includegraphics[width=0.7\columnwidth]{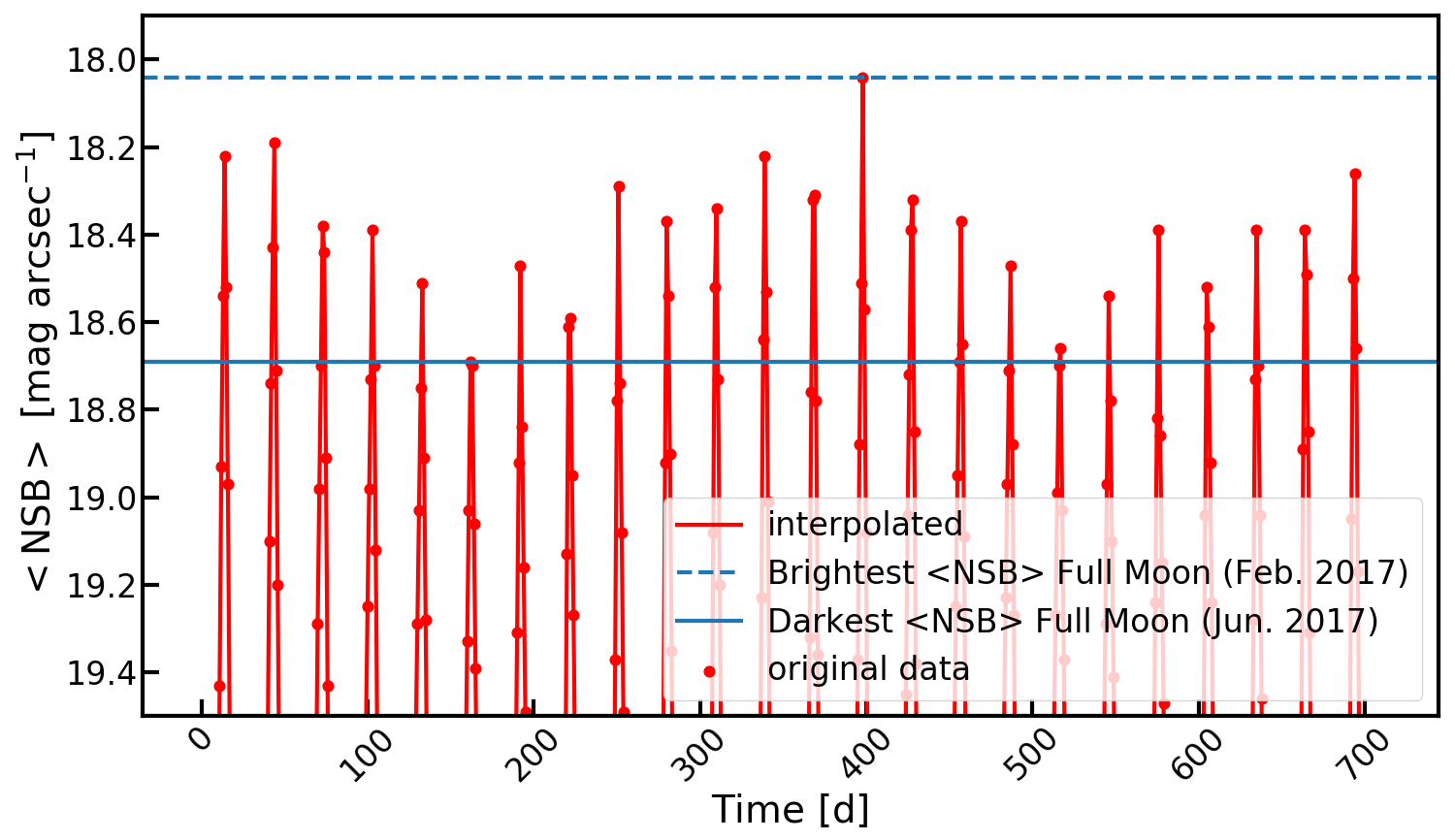}
        \caption[Skycalc Sky Model for one of our sites]{Skycalc sky model for the zenith, evaluated for one of our sites (VOE) for the years 2016--2017.
        The y-axis was limited to show only the bright peaks around full moon. Beside the circa-monthly oscillation, a seasonal
        variation is recognised, caused by changing altitude of the Moon between summer and winter.}
        \label{fig:SYNVOE}
        \end{center}
\end{figure}

Analogous to Section \ref{sec:appliillumodel}, we now evaluate the Skycalc model for one of our sites
(VOE) that we use as a proxy for our network. We do so for the
years 2016--2017.
However, in this case we calculate for each night the mean NSB (\NSB) rather than just the value at
midnight. The result is shown in Figure \ref{fig:SYNVOE}.
As for the illuminance model, a strong seasonal variation is recognized, that is mainly
caused by the changing altitude of the Moon. However, the relative change between brightest and darkest \NSB\
(18.04 vs. 18.69 $\mathsf{\magsqm}$) corresponds to a (linear-scale) factor of 1.8 only,
which is half the amplitude that is seen in the illuminance model. The main reason for this
discrepancy lies in the fact that we evaluate the illuminance model at midnight only,
i.e. at the point of maximum illumination, whereas for Skycalc we calculate the nightly means
within dark-time limits. Given the fact that summer nights are shorter, the relative contribution
of the full moon to the zenithal \NSB\ is thus higher in summer than in winter and the dynamic range
of the seasonal variation shrinks in that case.

%%%%%%%%%%%%%%%%%%%%%%%%%%%%%%%%%%%%%%%%%%%%%%%%%%%%%%%%%%%%%%%%%%%%%%%%%%%%%%%%%%%
%%%%%%%%%%%%%%%%%%%%%%%%%%%%%%%%%%%%%%%%%%%%%%%%%%%%%%%%%%%%%%%%%%%%%%%%%%%%%%%%%%%
%%%%%%%%%%%%%%%%%%%%%%%%%%%%%%%%%%%%%%%%%%%%%%%%%%%%%%%%%%%%%%%%%%%%%%%%%%%%%%%%%%%
%%%%%%%%%%%%%%%%%%%%%%%%%%%%%%%%%%%%%%%%%%%%%%%%%%%%%%%%%%%%%%%%%%%%%%%%%%%%%%%%%%%
%%%%%%%%%%%%%%%%%%%%%%%%%%%%%%%%%%%%%%%%%%%%%%%%%%%%%%%%%%%%%%%%%%%%%%%%%%%%%%%%%%%
% Data analysis (JP)
\section{Data analysis}

%%%%%%%%%%%%%%%%%%%%%%%%%%%%%%%%%%%%%%%%%%%%%%%%%%%%%%%%%%%%%%%%%%%%%%%%%%%%%%%%%%%
\subsection{Fourier analysis}\label{sec:fft}
Using our 2-year data of nightly mean \NSB\ values, we aim to reveal the amplitude of
the circalunar periodicity (and other periodic signals that might exist).
Numerous implementations of discrete Fourier transforms exist, but probably the
most common one is the Fast Fourier Transform (FFT), which we
also use for the analysis of our SQM data.
In particular, we apply the FFT algorithm as implemented
in \texttt{numpy} \citep{van2011numpy},
the fundamental package for scientific
computing with Python. In the following, we describe some general properties
of the FFT and highlight common obstacles of the method and how we treat them. 

\subsubsection{Which unit to choose?}\label{sec:fftunit}
The SQM delivers data in units of $\mathsf{\magsqm}$. However, conversions into linear units
are available, see Equation \ref{eq:unihedron}. Thus, we investigate how well
the amplitude of periodic variations such as the circalunar cycle
caused by varying moon phases or the seasonal rhythm,
that is driven by variations of the moon zenith distances between summer and winter, are
recovered in frequency space after application of the FFT in dependence of the data input unit.

For a linear input, the increase in zenithal night sky brightness from new moon to full moon starts very shallow,
almost unrecognizable. Then, few days before full moon, the increase accelerates to finally form a
sharp peak in the (time,\NSB) plane (see Figure \ref{fig:ifft_cdm}).
On the contrary, in logarithmic units such as $\mathsf{\magsqm}$ the phase of the shallow increase is
more pronounced (stretched in time) and the later steep increase somewhat compressed.
Thus, the apparent course in the
(time,\NSB) plane is rather smooth compared to linear units (see Figure \ref{fig:ifft}).

Since the FFT of a signal that is spread out in time delivers a compact result in frequency
space and vice versa,
logarithmic units such as $\mathsf{\magsqm}$ are preferred, because the smoother course
leads to a better definition
of the circalunar cycle in frequency space, i.e. a single peak.

\subsubsection{The role of gaps in \NSB\ measurements}\label{sec:fftgaps}
Gaps in SQM data series can occur due to several reasons,
be it hardware- or software-failures or even
meteorological conditions.
In this section, we investigate how gaps in the time series
affect the ability of the FFT algorithm to correctly reproduce
amplitudes of time-dependent variations. We do so by
introducing single and double gaps of varying lengths
into our synthetic sky model.
The recovered amplitudes of the circalunar and seasonal rhythm are then
evaluated against the gap fraction, i.e. the fraction of
data points on the equal-distant input time grid without measurement.

We stress that under all circumstances gaps should not be left as such,
but replaced with some form of interpolation or reasonable value. We
made tests using linear, quadratic and cubic spline interpolations, as well as
using the mean of the remaining data as fill value. We find that
the latter one is the most robust and preferred method. Although
cubic spline interpolation gives \textit{slightly} better results for
small gaps, i.e. when the gaps size is much smaller than the periodicity
of the desired signal, it can cause unforeseen results for larger gaps.

Our results are summarized in table \ref{tab:gaps}, which shows that
the single- and double
gap tests give very similar results:
gap fractions of 1\%, 5\%, 7.5\%
and 10\% recover the amplitude of the circalunar variation (A)
at levels of ~99\%, ~92--95\%, ~91\% and ~87--88\% respectively.
Even gap fractions
of 20\% recover ~75--77\% of the true amplitude. 

The seasonal variation (S) is less affected, because of its
longer periodic time. It can be accurately derived even for gap fractions of 20\%.

However, as described by \cite{Munteanu2016}, for even larger periods of missing data,
one should consider to perform spectral analysis using other algorithms
such as the \textit{Z transform} or the \textit{Lomb-Scargle algorithm}.
They might be able to reproduce the amplitude for cases where gaps
make up more than 50 percent of the time series.

Additionally, the presence of gaps in the time series leads
to an increase of `frequency noise' in the amplitude spectrum,
limiting the chance to detect low-amplitude variations
at certain frequencies. As shown in
Table \ref{tab:gaps}, the noise roughly doubles between 0 and 5\% gap fractions,
but stays almost constant from thereon.

\begin{table}
\centering
\caption[Influence of gaps on the FFT amplitude spectrum]{We test the influence of
single and double gaps on the ability to recover signal amplitudes
from synthetic sky model data. The gap fraction in percent is given in \textit{column 1}
and the recovered amplitudes of the circalunar and the seasonal variation are
shown in \textit{column 2 and 3} for the single gap case, and \textit{column 5 and 6}
for the double gap test. FFT Noise is given in \textit{column 4 and 7}. The unit
for all measurements is $\mathsf{\magsqm}$.}
\label{tab:gaps}
\begin{tabular}{lccccccc}
\multicolumn{1}{c}{} & \multicolumn{3}{c}{single gap} &  & \multicolumn{3}{c}{double gap} \\
\multicolumn{1}{c}{GF} & A & S & N &  & A & S & N \\
\multicolumn{1}{c}{(1)} & (2) & (3) & (4) &  & (5) & (6) & (7) \\
\hline \hline
0 & 1.508 & 0.073 & 0.006 &  & 1.508 & 0.073 & 0.006 \\
1 & 1.489 & 0.077 & 0.008 &  & 1.488 & 0.075 & 0.009 \\
5 & 1.415 & 0.078 & 0.013 &  & 1.438 & 0.083 & 0.011 \\
7.5 & 1.397 & 0.070 & 0.012 &  & 1.403 & 0.078 & 0.014 \\
10 & 1.356 & 0.081 & 0.012 &  & 1.349 & 0.075 & 0.015 \\
20 & 1.217 & 0.072 & 0.014 &  & 1.205 & 0.074 & 0.015 \\
30 & 1.037 & 0.060 & 0.015 &  & 1.067 & 0.067 & 0.015 \\
50 & 0.754 & 0.036 & 0.013 &  & 0.765 & 0.010 & 0.015
\end{tabular}
\end{table}

\subsubsection{Importance of the length of the data series for FFTs}\label{sec:fftdatalength}
The length of a time series as well as its sampling rate are of
importance for FFT studies, because they define the frequency
resolution in the final amplitude spectrum.

Our sampling rate $\mathsf{f_s}$ is one measurement per night or 
$\mathsf{f_s=\,1\,d^{-1}}$. The distance $\mathsf{\Updelta t}$ between two data points is:
$\mathsf{\Updelta t\,=\,\frac{1}{f_s}\,=\,1\,d}$.
The final frequency range is thus limited to the interval $\mathsf{[-\frac{f_s}{2};+\frac{f_s}{2}]}$,
i.e. $\mathsf{-0.5...0.5\,d^{-1}}$. The highest measurable frequency is
$\mathsf{0.5\,d^{-1}}$ or two days periodic time.
The number of discrete points $\mathsf{N}$ in the final frequency domain equals
the number in time domain. Thus, the distance $\mathsf{\Updelta \nu}$ in frequency space is 
$\mathsf{\Updelta \nu\,=\,\frac{f_s}{N}}$, which shows that the
frequency resolution is controlled by the sampling rate $\mathsf{f_s}$
and the length of the time series $\mathsf{N}$. Since our sampling rate is fixed,
the number of data points is the main quantity that drives the frequency resolution
in our final amplitude spectrum.

For example, if we wanted to \textit{detect} the frequency of the lunar synodic month
(without prior knowledge), with an accuracy of 0.5\,d, we would need at least 1711
data points, equivalent to 4.68 years at a sampling rate of one
measurement per night.

However, with prior knowledge of the period - in our case 29.5\,d for the synodic month -
one can adjust the time axis such that the final discrete amplitude spectrum covers the
corresponding frequency, i.e. $\mathsf{\frac{1}{29.5}\,=\,0.0339\,d^{-1}}$. This is
achieved by limiting the data points such that no discontinuities occur at the
edges of the time series, i.e. spectral leakage (see next paragraph) is eliminated.
That way, even time series of only one year recover the amplitude on levels better than
90 percent.

\subsubsection{Avoiding spectral leakage}\label{sec:spectralleakage}
The ability of the FFT algorithm (and any other discrete Fourier transforms)
to recover amplitudes is limited due to the fact that the duration of the observation
is finite. This means that the input signal is factual a product
with a rectangular window. The discrete spectrum of any
finite signal is thus spread out over multiple frequency components and
the amplitude is not fully recovered anymore. This behaviour is called \textit{spectral leakage}.

However, the effect may be reduced by 1) avoiding discontinuities of the input
signal or 2) gradually decreasing the amplitude of the signal towards the 
edges of the measurement series. The first method requires prior knowledge of
the periodic time and phase of the signal of interest (which in most applications is not
fulfilled). The latter case can be achieved by multiplying the input time
series with a window function, e.g. a Hanning window,
before the FFT is applied.

Since the periodicity of the synodic month is known,
we could test both scenarios using sinusoidal and our synthetic sky
time series as input. Although the application of a Hanning window improved
amplitude recovery in cases where discontinuities occurred at the edges,
we find that an continuous input time series
(e.g. from first new moon 2016 to
last new moon in 2017) gives best results with only negligible amounts of spectral
leakage and amplitude recovery at levels better than 95 percent.
Thus, for our main science case, i.e. studying the
circalunar rhythm, a continuous time series as input for the FFT is recommended
and all our FFT analysis was performed using data from new moon 2016/01/09 to new moon 2017/12/18 only.

%%%%%%%%%%%%%%%%%%%%%%%%%%%%%%%%%%%%%%%%%%%%%%%%%%%%%%%%%%%%%%%%%%%%%%%%%%%%%%%%%%%
%%%%%%%%%%%%%%%%%%%%%%%%%%%%%%%%%%%%%%%%%%%%%%%%%%%%%%%%%%%%%%%%%%%%%%%%%%%%%%%%%%%
%%%%%%%%%%%%%%%%%%%%%%%%%%%%%%%%%%%%%%%%%%%%%%%%%%%%%%%%%%%%%%%%%%%%%%%%%%%%%%%%%%%
%%%%%%%%%%%%%%%%%%%%%%%%%%%%%%%%%%%%%%%%%%%%%%%%%%%%%%%%%%%%%%%%%%%%%%%%%%%%%%%%%%%
%%%%%%%%%%%%%%%%%%%%%%%%%%%%%%%%%%%%%%%%%%%%%%%%%%%%%%%%%%%%%%%%%%%%%%%%%%%%%%%%%%%
%%%%%%%%%%%%%%%%%%%%%%%%%%%%%%%%%%%%%%%%%%%%%%%%%%%%%%%%%%%%%%%%%%%%%%%%%%%%%%%%%%%
% Results and discussion
\section{Results}

%%%%%%%%%%%%%%%%%%%%%%%%%%%%%%%%%%%%%%%%%%%%%%%%%%%%%%%%%%%%%%%%%%%%%%%%%%%%%%%%%%%
\subsection{FFT analysis of the illuminance and Skycalc models}\label{sec:fftmodel}
\begin{figure}
        \begin{center}
        \includegraphics[width=0.7\columnwidth]{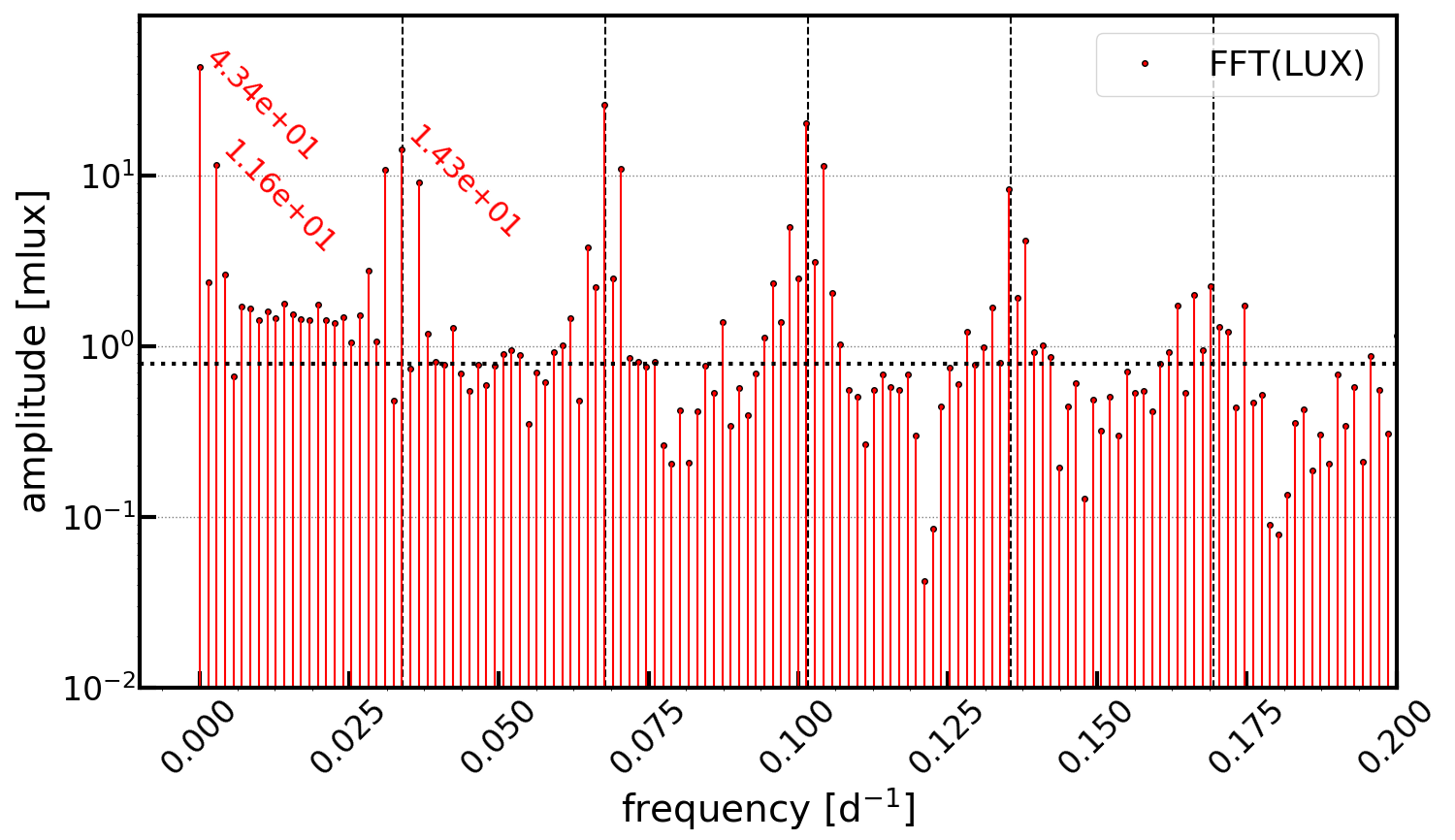}
        \caption[FFT Analysis of the Illuminance Model]{FFT amplitude spectrum of the illuminance model as described in Section \ref{sec:illuminancemodel}. The shown frequency range is limited to values between 0 and 0.2. Note that the unit of the
        y-axis is \textit{milli-lux}. Labeled amplitudes (\textit{from left to right} correspond to the mean of all input data, the seasonal variation and the circalunar cycle.}
        \label{fig:LUXVOEfft}
        \end{center}
\end{figure}

\begin{figure}
        \begin{center}
        \includegraphics[width=0.7\columnwidth]{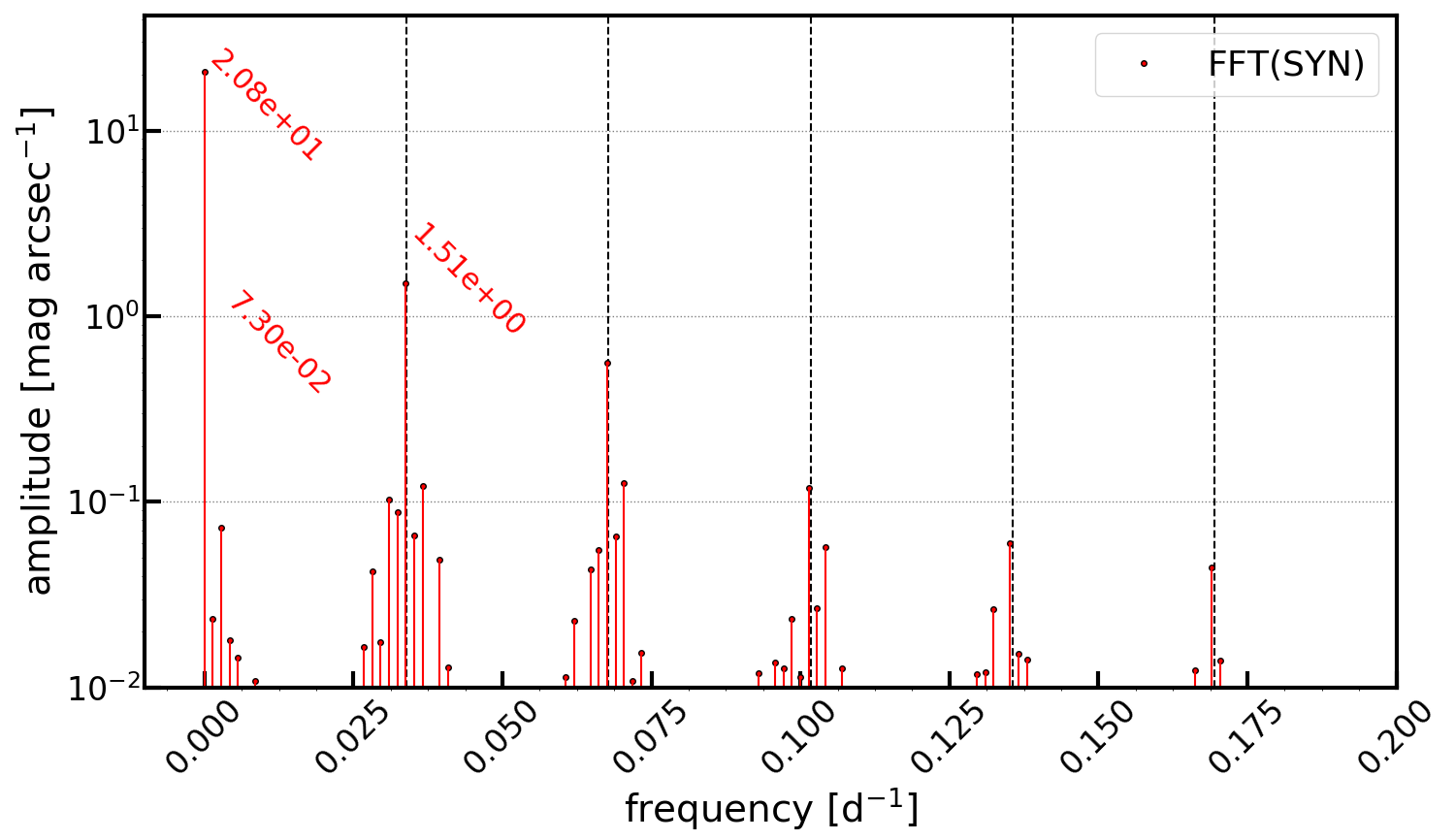}
        \caption[FFT Analysis of the Skycalc Sky Model]{FFT amplitude spectrum of the Skycalc sky model as described in Section \ref{sec:skymodel}. The shown frequency range is limited to values between 0 and 0.2.
        Labeled amplitudes (\textit{from left to right} correspond to the mean \NSB\ of the input data,
        the seasonal variation and the circalunar cycle.}
        \label{fig:SYNVOEfft}
        \end{center}
\end{figure}

As a reference for our measurement sites, we perform an FFT
analysis of our model data using the considerations from Section \ref{sec:fft}.
To do so, for the Skycalc model we first calculate the nightly averages, i.e.
\NSB, as described in Section \ref{sec:locmeth}.
The resulting FFT amplitude spectra are shown in Figures \ref{fig:LUXVOEfft} and
\ref{fig:SYNVOEfft}.

An analysis of both models reveals significantly strong amplitudes at
the frequency of $\mathsf{\frac{1}{29.5}\,=\,0.0339\,d^{-1}}$,
i.e. the fundamental of the circalunar rhythm. Also its harmonics are identified
at multiples of that frequency.
However, the noise level in the illuminance model is higher
because of its peak-like input (see Section \ref{sec:fftunit}).

We also note that the zero-frequency amplitude is not comparable between the models.
While for the Skycalc model, the zero-frequency is in fact a representation of the
mean \NSB, this is not true for the illuminance model, because for the FFT
analysis all input data with illuminance levels of 0lux is considered as gaps and
are thus filled with the mean in order to improve recognition of the
circalunar rhythm as explained in Section \ref{sec:fftgaps}.

It is also recognized that the seasonal variation seen in the amplitude spectrum of the
illuminance model is stronger than in the Skycalc model.
This is caused because we only consider midnight values for the illuminance model,
while for the Skycalc model nightly averages are calculated.
Hence, the contribution of the peak NSB at midnight is smoothed out in time due to averaging,
while it is fully captured for midnight data.
The averaging effect is even stronger in winter when the Moon's
contribution to the zenithal \NSB\ is highest,
because then the nights are longer.
The seasonal signal is thus equalized throughout the year.

%Warum Modelle...

%Ground Illumination berücksichtigt keine anderen Faktoren außer Mond, natürlicher Himmel wäre +1 mlx

%Referenz auf Illuminance Model aus Vöcklabruck? Dann sollten wir bei der Caption Section 3.1.2 bzw. 3.2.5 schreiben...

%Beschreibung: zu sehen ist nur Einfluss des Mondes, eindeutige Frequenzen hier sichtbar, besonders die seasonal variations

%Beide Modelle zeigen die gleichen Peaks, aber mit unterschiedlichen Stärken, Illuminance model zeigt viele Frequenzen mehr, lower S/N, da nur ein Zeitpunkt pro Nacht gewählt wurde anstatt <NSB> ? Wie bereits in 3.2.5 angemerkt

%Modelle sind einzige Möglichkeit nur den Einfluss des Mondes zu bestimmen, wenngleich nur theoretisch, nur ohne Einfluss von Lichtverschmutzung

%%%%%%%%%%%%%%%%%%%%%%%%%%%%%%%%%%%%%%%%%%%%%%%%%%%%%%%%%%%%%%%%%%%%%%%%%%%%%%%%%%%
\subsection{Identifying the synodical month in the \NSB\ data and quantifying its amplitude}
\begin{figure*}
        \begin{center}
        \includegraphics[width=\textwidth]{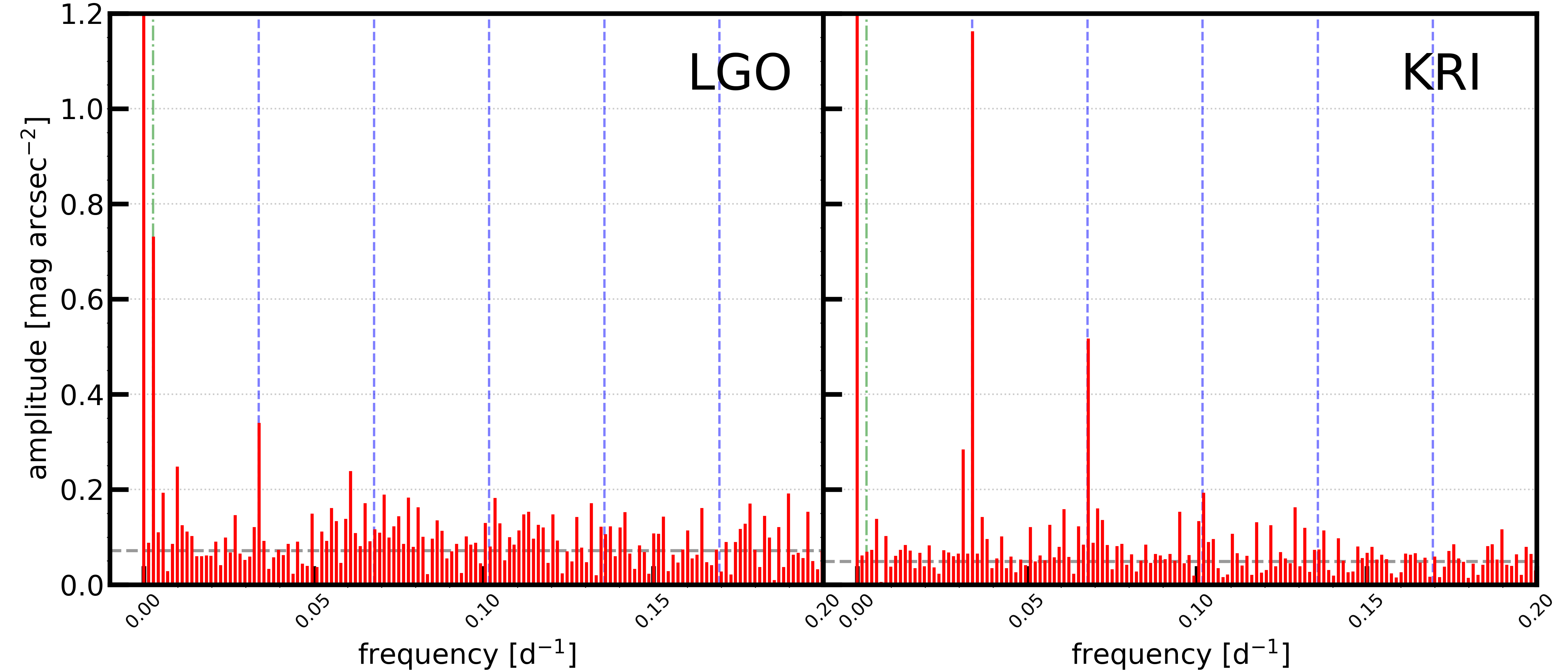}
        \caption[FFT of an urban and rural site]{FFT amplitude spectrum of an urban (LGO) and rural (KRI) site. The shown frequency range is limited to values between 0 and 0.2 and
        the amplitude range is cut at 1.2 $\mathsf{\magsqm}$ in order to focus on amplitudes
        in the given frequency range, but excluding the peak at zero, i.e the mean
        nightsky brightness.}
        \label{fig:lgo_vs_kri}
        \end{center}
\end{figure*}

%Zunächst unser Hauptziel gewesen, erklären, dass und in welcher Form erreicht
%Beispiel-Abbildungen: sehr naturnaher versus sehr lichtverschmutzter Standort, zB LOS oder KRI versus LGO oder LSM
From our SQM data, we first calculate the nightly averages, i.e. \NSB, 
as described in section \ref{sec:locmeth} and then perform an FFT analysis using
the considerations from section \ref{sec:fft}.
That way, we can clearly detect the circalunar rhythm at all our sites (see Table \ref{tab:fftresult}), 
be it rural or urban. However, the amplitude decreases from 1.55 to 0.33\,$\mathsf{\magsqm}$,
corresponding to a factor of three on a linear scale, from the darkest to
the brightest sites. A comparison between amplitude
spectra of a typical urban and rural site is shown in Figure \ref{fig:lgo_vs_kri}.

\begin{figure}
        \includegraphics[width=0.5\columnwidth]{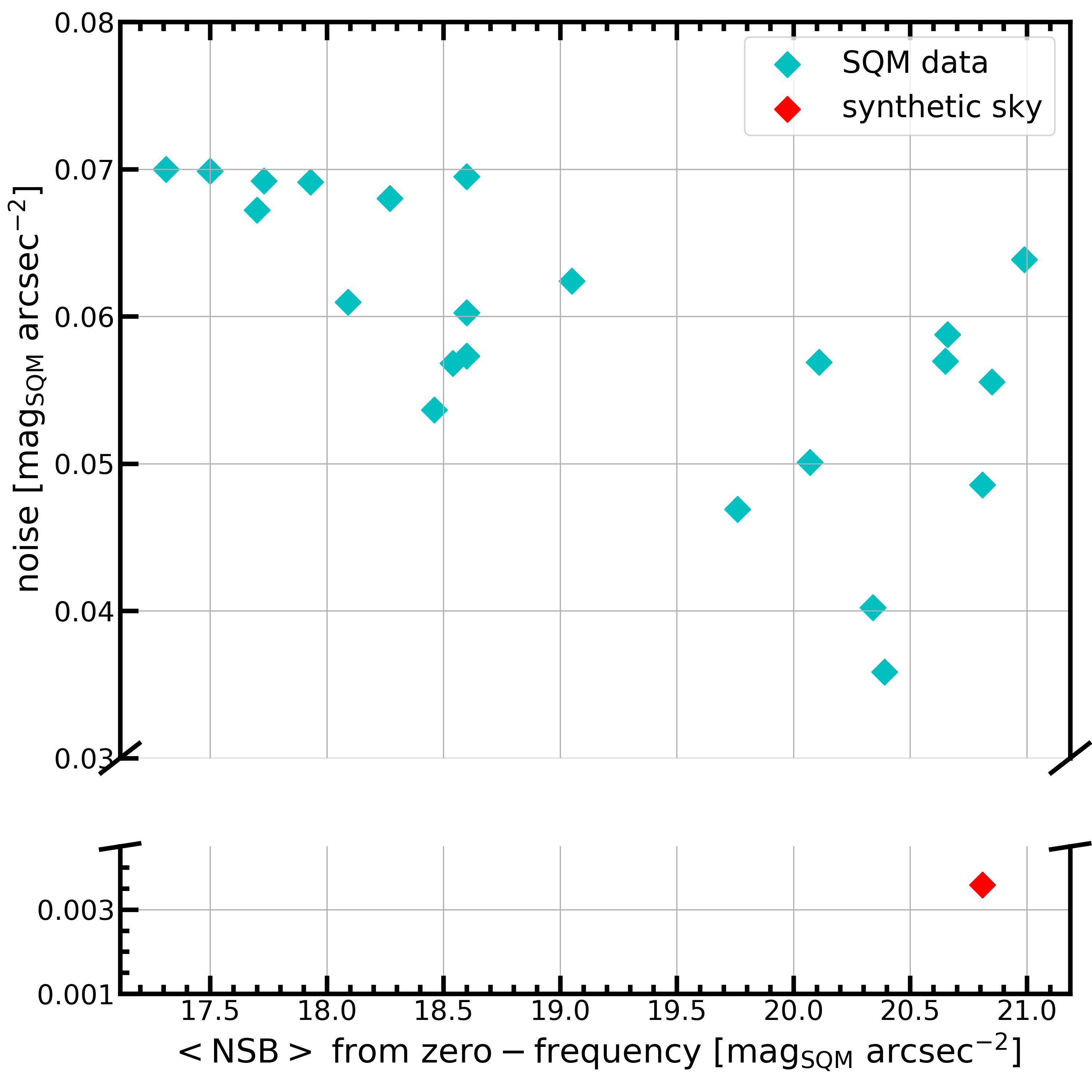}
        \caption[FFT noise]{FFT noise increases with the level of light pollution.}
        \label{fig:fftnoise}
\end{figure}

It is also recognized that at rural sites, harmonics of the main
variation at a frequency $\mathsf{0.0339\,d^{-1}}$ can be identified
up to 3rd order. In contrast, at urban sites, they perish in frequency
noise, which increases with light pollution.
In fact, the FFT noise roughly shows a bimodal distribution between rural and
urban sites, as seen in Figure \ref{fig:fftnoise}.
This is, because day-to-day variations are more pronounced at urban, light-polluted sites due to
backscattering of ALAN at clouds
\citep{Kyba2011,Kyba2012,Puschnig2014a,Puschnig2014b,Aube2016}.
Thus, the observed night sky brightness
typically jumps between two preferred values; see e.g. \cite[Figures A1a, A1b and 7]{Posch2018}.
Since the noise as calculated here
is dominated by high frequencies, i.e. roughly 64 percent of the frequencies
correspond to periodic times of equal or less than five days, the same
bimodality is seen here.

Another interesting feature seen in Figure \ref{fig:fftnoise} is that the scatter of the noise level
increases along the \NSB\ axis, i.e. darker sites show larger variance in the noise. The cause of
this trend might be explained by findings of \cite{Kocifaj2014}, who showed that the
amplification factor due to clouds decreases with increasing city size and thus the level
of light pollution.

\begin{table}
\centering
\caption[Summary of recognized features in the FFT amplitude spectrum]{Summary of
recognized features in the FFT amplitude spectrum for 23 SQM stations and the synthetic sky model (SYN).
\textit{Column 1} is the station code, \textit{column 2 and 3}, the average night sky brightness from
the peak at zero frequency in units of $\mathsf{10^{-5}\,\watts}$\ and $\mathsf{\magsqm}$
respectively. The amplitude of the circalunar cycle in $\mathsf{\magsqm}$ is shown in \textit{column 4}
and the circalunar contrast CLC,
i.e. the amplitude expressed in percent of $\mathsf{\NSB}$, is found in \textit{column 5}.
The amplitude of the seasonal variation (bright winters, dark summers in units of
$\mathsf{\magsqm}$ is given in column \textit{column 6} and the noise in $\mathsf{\magsqm}$
in the \textit{last column}.}
\label{tab:fftresult}
\begin{tabular}{ccccccc}
COD & $\mathsf{\NSB_W}$ & $\mathsf{\NSB_{mag}}$ & A & CLC & S & N \\
(1) & (2) & (3) & (4) & (5) & (6) & (7) \\
\hline \hline \\
LGO & 18.25 & 17.31 & 0.34 & 36.8 & 0.73 & 0.071 \\
LSM & 15.32 & 17.51 & 0.33 & 34.3 & 0.77 & 0.070 \\
WEL & 12.74 & 17.70 & 0.36 & 39.3 & 0.86 & 0.067 \\
LSW & 12.39 & 17.73 & 0.41 & 45.9 & 0.58 & 0.069 \\
STW & 10.31 & 17.93 & 0.45 & 50.0 & 0.49 & 0.069 \\
TRA & 8.90 & 18.09 & 0.46 & 52.8 & 0.55 & 0.061 \\
STY & 7.54 & 18.27 & 0.50 & 58.5 & 0.69 & 0.068 \\
PAS & 6.33 & 18.46 & 0.53 & 62.9 & 0.57 & 0.054 \\
BRA & 5.88 & 18.54 & 0.66 & 83.7 & 0.67 & 0.057 \\
GRI & 5.56 & 18.60 & 0.57 & 69.0 & 0.92 & 0.057 \\
VOE & 5.56 & 18.62 & 0.64 & 82.0 & 0.74 & 0.060 \\
FRE & 5.56 & 18.60 & 0.71 & 92.3 & 0.76 & 0.070 \\
MAT & 3.67 & 19.05 & 0.83 & 114.8 & 0.65 & 0.062 \\
MUN & 1.91 & 19.76 & 1.10 & 175.4 & 0.56 & 0.047 \\
GIS & 1.44 & 20.07 & 1.02 & 155.9 & 0.06 & 0.050 \\
ULI & 1.38 & 20.13 & 1.16 & 191.1 & 0.50 & 0.057 \\
KID & 1.12 & 20.37 & 1.19 & 204.8 & 0.09 & 0.040 \\
GRU & 1.07 & 20.39 & 1.34 & 243.6 & 0.38 & 0.036 \\
FEU & 0.84 & 20.66 & 1.22 & 207.6 & 0.07 & 0.057 \\
LOS & 0.83 & 20.86 & 1.30 & 243.6 & 0.01 & 0.059 \\
KRI & 0.73 & 20.81 & 1.16 & 191.1 & 0.07 & 0.049 \\
ZOE & 0.70 & 21.01 & 1.55 & 324.6 & 0.10 & 0.056 \\
BOD & 0.62 & 21.04 & 1.51 & 301.8 & 0.10 & 0.064 \\
SYN & 0.73 & 20.81 & 1.51 & 301.8 & 0.08 & 0.004
\end{tabular}
\end{table}

%%%%%%%%%%%%%%%%%%%%%%%%%%%%%%%%%%%%%%%%%%%%%%%%%%%%%%%%%%%%%%%%%%%%%%%%%%%%%%%%%%%
\subsection{Relation between the circalunar amplitude and zenithal \NSB\ for mid latitudes}
\begin{figure*}
        \includegraphics[width=\textwidth]{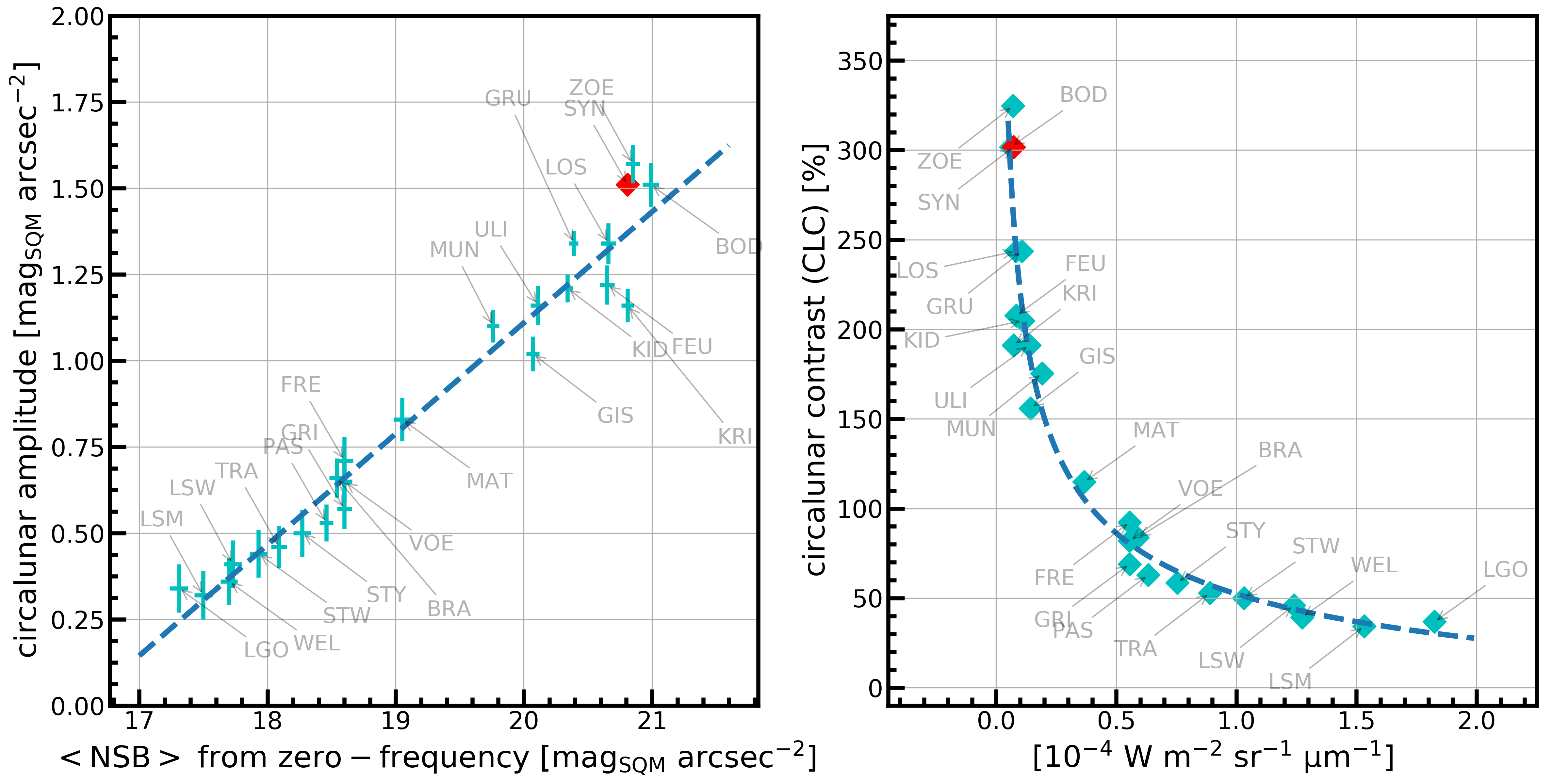}
        \caption[Relation between the circalunar amplitude and zenithal \NSB\ for mid latitudes]
        {\textit{Left panel}: Relation between the circalunar amplitude and zenithal \NSB\
        for mid latitudes ($\sim48^\degree$). Individual points denote amplitudes derived
        through FFT analysis, with $\pm 1\upsigma$ errors found from the noise in the FFT amplitude spectra.
        The data is in good agreement with a linear relation with a scatter of only 0.06 $\mathsf{\magsqm}$.
        The red point denotes amplitude and \NSB\ for our synthetic Skycalc sky model. 
        \textit{Right panel}: The same relation as shown in the left panel, but converted to luminance
        using Equation \ref{eq:unihedron} and normalized to the moonless, zenithal luminance of the synthetic sky model,
        i.e. the contrast of the Moon's luminance at zenith against skyglow.}
        \label{fig:circalunarrelation}
\end{figure*}

Previous observations have already qualitatively shown, that the circalunar rhythm
steadily fades away with an increasing level of anthropogenic light at night.
Using our FFT analysis, we can now quantify how ALAN affects the lunar cycle's
degree of recognition. We do so by plotting the mean nightsky brightness,
given by the amplitude at zero-frequency, against the amplitude of the circalunar rhythm.
The result is shown in the left panel of Figure \ref{fig:circalunarrelation}. A linear fit (see Equation
\ref{equ:nsbamprelation}) is found to be appropriate, with a scatter of only 0.062 $\mathsf{\magsqm}$.
Extrapolation leads us to an \NSB\ level of 16.5\,$\mathsf{\magsqm}$, the level
at which the circalunar rhythm practically vanishes (at zenith) and is indistinguishable
from the polluted nightsky.
\begin{equation}\label{equ:nsbamprelation}
\mathsf{
    A\,=\,0.322\,\NSB\,-\,5.324
}
\end{equation}
We may also convert the amplitudes in linear units, e.g. to luminance (see Equation \ref{eq:unihedron}),
and then express
the amplitude of the circalunar cycle as multiple of \NSB, which we denote as
\textit{circalunar contrast} (CLC). As shown in the right panel of
Figure \ref{fig:circalunarrelation}, the contrast between the average
\NSB\ level and the darkening/brightening due to new/full moon ranges between
30 and 300 percent for our urban and rural sites respectively.

%%%%%%%%%%%%%%%%%%%%%%%%%%%%%%%%%%%%%%%%%%%%%%%%%%%%%%%%%%%%%%%%%%%%%%%%%%%%%%%%%%%
%%%%%%%%%%%%%%%%%%%%%%%%%%%%%%%%%%%%%%%%%%%%%%%%%%%%%%%%%%%%%%%%%%%%%%%%%%%%%%%%%%%
\subsection{Searching for other than lunar frequencies in the \NSB\ data}
As seen in Sections \ref{sec:models} and \ref{sec:fftmodel} a naturally occuring seasonal variation
caused by the changing height of the ecliptic (and thus the Moon)
is expected. In our measurements, however, for the darkest, rural stations (FEU, KRI, LOS, ZOE, BOD),
we do not recover any such seasonal variation at a significant level (compare Table \ref{tab:fftpeaks}).
This is mainly because our \NSB\ data includes overcast skies that weaken the amplitude of
the signal in an unforeseeable way, and given the fact that we cover only two years, the signal easily
vanishes.

On the other side, we do detect a very strong seasonal variation at all urban and also at most
intermediate stations (compare Table \ref{tab:fftpeaks}). However, the amplitude of that signal is too
large to be caused by the Moon. This oscillation might be the result of climatological
effects that enhance ALAN during winters, i.e. combined effects of increased surface albedo and
lower vegetation state \citep{Wallner2019}.
The observed seasonal variation might also be related to the aerosol optical depth (AOD).
As demonstrated by \cite{Aube2015}, the zenith radiance can
increase several tens of times when optical depth is significantly lowered.
However, disentangling the contribution of the several effects (surface albedo, vegetation, AOD)
is beyond the scope of this paper and would require ancillary data products.

At urban sites (LGO, LSM, WEL, LSW, STY, PAS, BRA, GRI, VOE, MAT), we further detect a
significant ($>$3 sigma) signal with a periodic time of $\sim$100 days. The cause of this roughly quaterly
variation is unclear and was not reported previously. In our Table \ref{tab:fftpeaks} we denote it
as \textit{unknown}.

We also search for weekly variations that might be associated
with an increased human nighttime activity on weekends. Although we do see at our urban
stations amplitudes on the order of 0.15--0.20 $\mathsf{\magsqm}$, these are not significant
($<$3 sigma). Furthermore, its frequency is very close to that of the third lunar harmonic.

\begin{figure}
        \includegraphics[width=0.5\columnwidth]{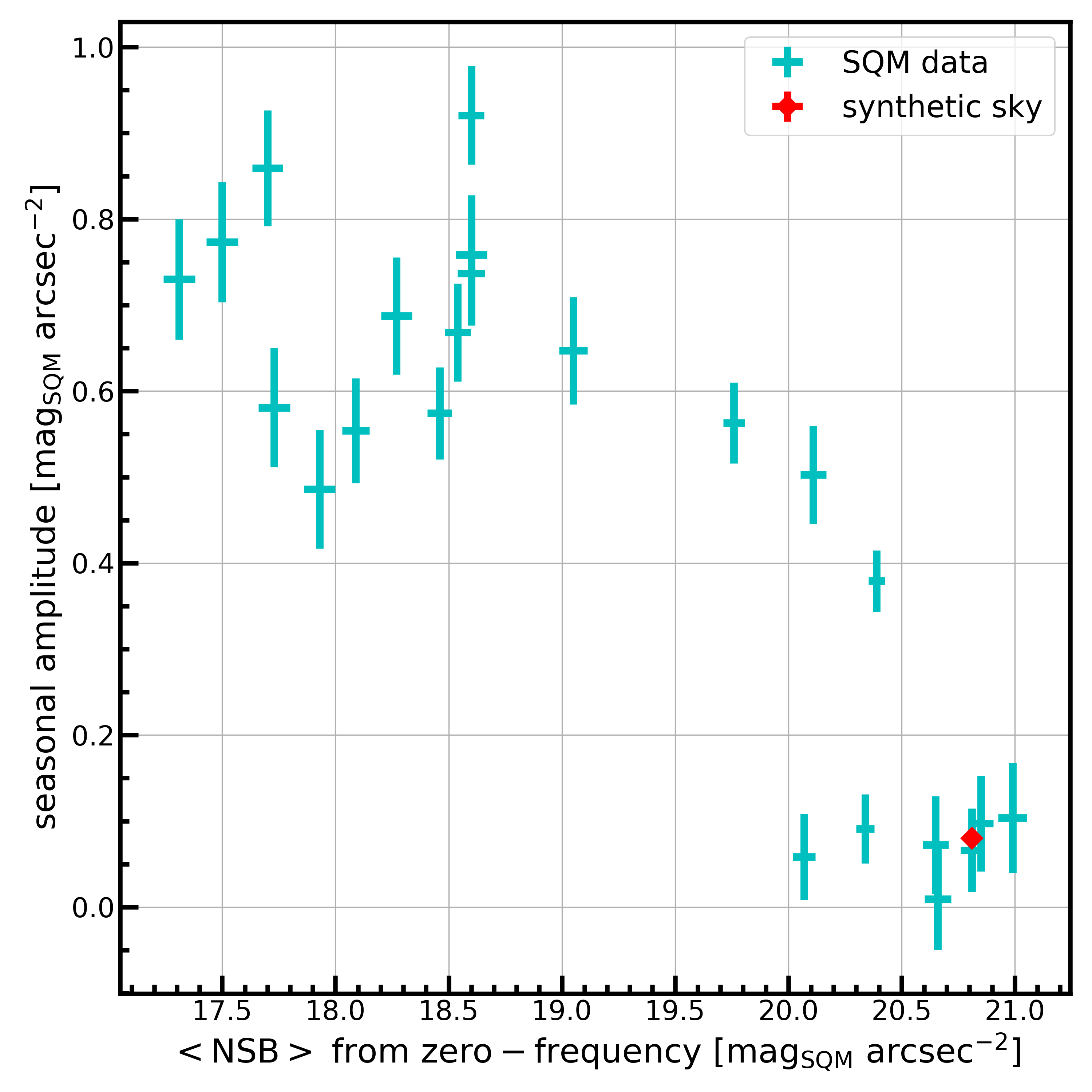}
        \caption[Seasonal relation]{The amplitude of the seasonal oscillation against \NSB,
        both derived through FFT analysis of nightly \NSB\ values. Each point corresponds to
        one of our SQM stations, with error bars indicating the $\pm 1\upsigma$ errors found from
        the noise in the FFT amplitude spectrum. The red point shows the location of our
        synthetic Skycalc sky model, indicating that for our SQM network the amplitude of the
        naturally occuring seasonal variation due to the Moon (its seasonal change of altitude)
        has an amplitude of less than 0.1$\mathsf{\magsqm}$, which is too low to be detectable
        within the frequency noise as we see in our FFT amplitude spectra. Thus, no clear trend
        or relation is identified. The strong seasonal amplitudes seen at urban sites are thus
        caused by other seasonal effects such as changing albedo or vegetation state between
        winter and summer (Puschnig et al. in prep.).
        }
        \label{fig:seasonalrelation}
\end{figure}

%%%%%%%%%%%%%%%%%%%%%%%%%%%%%%%%%%%%%%%%%%%%%%%%%%%%%%%%%%%%%%%%%%%%%%%%%%%%%%%%%%%
%%%%%%%%%%%%%%%%%%%%%%%%%%%%%%%%%%%%%%%%%%%%%%%%%%%%%%%%%%%%%%%%%%%%%%%%%%%%%%%%%%%
%%%%%%%%%%%%%%%%%%%%%%%%%%%%%%%%%%%%%%%%%%%%%%%%%%%%%%%%%%%%%%%%%%%%%%%%%%%%%%%%%%%
%%%%%%%%%%%%%%%%%%%%%%%%%%%%%%%%%%%%%%%%%%%%%%%%%%%%%%%%%%%%%%%%%%%%%%%%%%%%%%%%%%%
%%%%%%%%%%%%%%%%%%%%%%%%%%%%%%%%%%%%%%%%%%%%%%%%%%%%%%%%%%%%%%%%%%%%%%%%%%%%%%%%%%%
\section{Discussion and Conclusion}
Prior to a discussion of possible limitations of our method and implications
of our findings in a wider context,
we first compare our results to those of
\cite{Bara2016}, who performed an SQM based FFT analysis of 14 Galician
stations using a 1-year dataset.
\cite{Bara2016} have previously defined the \textit{moonlight factor} (y)
as the ratio of \textit{power densities} between the fundamental circalunar frequency and the
zero frequency. Note that they used double-sided power density spectra, while we use single-sided
amplitude spectra. The power spectra of \cite{Bara2016} thus show negative and positive
frequency components $k>0$ with heights of $\sfrac{A_k^2}{4}$ compared to our amplitudes (A)
in Table \ref{tab:fftresult} and at a height of $A_0^2$ for the
zero-frequency component ($\mathsf{\NSB_{mag}}$).
They reported ranges for the moonlight factor between 0.2--0.3$\times10^{-3}$ for urban sites,
1.5--2.4$\times10^{-3}$ for dark rural sites and values in-between for intermediate
regions. A compilation of the moonlight factor calculated for our sites is shown in Table
\ref{tab:moonlightfactor}.
We find that the moonlight factor derived through our FFT methodology and data
is almost a factor of 2 lower
than those reported in \cite{Bara2016}. Although it is expected that the Moon's impact on zenithal NSBs
in Galicia is stronger than in Upper Austria, because the Galician network's
geographic latitude is lower by $\sim$5$^{\degree}$ and thus the ecliptic and Moon closer to zenith,
a factor of 2 seems to be relatively high.
We argue that the relatively large difference is mainly caused because
\cite{Bara2016} take into account NSB measurements obtained at
midnight only rather than averaging over dark time hours as we do.
As a result -- in analogy to the explanation in Section \ref{sec:fftmodel} --
for nightly averages peak values that occur at midnight are smoothed in time,
which is also the reason why we do not capture the seasonal variation
(compare Figure \ref{fig:seasonalrelation}).
On the other hand, averaging over dark times has the advantage of providing
a better definition and thus less scatter of the amplitudes over several months, because a single
nightly measurement may easily be affected by short-term perturbations
such as cloud cover, which may impact NSB measurements in a complex way:
Considering that backscattering of moonlight on scattered clouds (in zenith) exists in analogy
to previous findings of \cite{Kyba2011,Kyba2012,Puschnig2014a,Puschnig2014b},
clouds may enhance the zenithal NSB on the one hand, but it may also lower the 
zenithal NSB on the other hand, e.g for fully overcast skies,
similar to previous findings of \citep{Posch2018,Jechow2019}.

However, recognition of the circalunar amplitude as performed by \cite{Bara2016}
is affected by spectral leakage that lowers the amplitude (compared to our approach)
due to smearing out of the signal over multiple frequencies, as explained in
Section \ref{sec:spectralleakage}. Additionally, \cite{Bara2016}
use only 1-year of input data, leading to lower frequency resolution and thus even
more leakage, as explained in Section \ref{sec:fftdatalength}.

We conclude that a combination of these effects causes the factor 2 discrepancy, but
we stress that the linear relation in Figure \ref{fig:circalunarrelation},  its intercept mainly,
would not change by a factor 2 for geographic locations similar to those of the Galician
network. This is shown in the following.

In principle, it is expected that the relation in Figure \ref{fig:circalunarrelation} shifts
towards lower circalunar amplitudes for geographic latitudes north of 48$^\degree$ and
towards higher amplitudes otherwise.
In order to asses how much the relation's intercept may shift, we perform
FFT analysis of 2-year Skycalc models calculated for different latitudes.
For the first test, we choose a latitude of N 36$^{\degree}$ (e.g. Gibraltar).
The derived amplitude of the circalunar cycle is 1.61$\mathsf{\magsqm}$,
which is only $\sim$0.1$\mathsf{\magsqm}$
higher than for the synthetic models calculated for our network.
For the second test, we choose a latitude of N 60$^{\degree}$ (e.g. slightly north of Stockholm).
However, we caution that for latitudes above 51.5$^{\degree}$,
during summer there are no dark times with the sun being more than 15$^{\degree}$ below the horizon.
As a result, gaps in the input data limit the recoverability of the circalunar amplitude,
as explained in Section \ref{sec:fftgaps}. In that case we thus find
an amplitude of 1.02$\mathsf{\magsqm}$ only, which is mainly due to the large summer gaps accounting
for roughly 25 percent of the input data.
From these tests we conclude that our relation between the circalunar amplitude and \NSB\
is at least valid for latitudes between $\sim$40--50$^\degree$.

Our linear relation between zenithal \NSB\ and the circalunar amplitude in Figure \ref{fig:circalunarrelation}
implies that the circa-monthly variation of moonlight is still traceable over large areas, not only in Upper
Austria, but also in many other countries (compare light pollution atlas by \cite{Falchi2016}).
The circalunar rhythm is thus expected to practically vanish due to ALAN only in the innermost parts of
major cities where the zenithal NSB may exceed 16.5 $\mathsf{\magsqm}$.
However, at the same time, it is recognized that the contribution
of the Moon to the total zenithal NSB (i.e. contrast) is largely reduced due to ALAN,
namely by a factor of $\sim\sfrac{1}{9}$ for
urban areas (e.g. Linz with $\sim$200,000 inhabitants), a factor of $\sim\sfrac{1}{3}$ for small towns with
less than 10,000 inhabitants (e.g. Freistadt or Mattighofen) and still
up to $\sim\sfrac{1}{2}$ for one of our rural stations (e.g. M\"unzkirchen, a village with less than 2,600
inhabitants). Only two of our sites, both situated in national parks (Bodinggraben and Z\"oblboden), show
natural circalunar amplitudes.

Finally, we discuss implications of the SQM's spectral bandpass
that covers a range of $\sim$300--680$\upmu$m. Although this is very close to the so called
\textit{photosynthetically active radiation}, i.e. the spectral range between 400--700$\upmu$m
to which photosynthetic organisms are sensitive, we caution that within that spectral range, Chlorophyll
-- the most abundant plant pigment -- has a sensitivity curve that is very different from the
SQM's sensitivity curve shown in Figure \ref{fig:sqmtrans}.
Rather than being mostly sensitive to green photons, Chlorophyll is mostly sensitive to red and blue photons.
The exact degradation of the circalunar rhythm as recognized by photosynthetic organisms
may thus be different from what we observe using SQMs.
However, other species such as e.g. ocean fish, that have maximum spectral sensitivities between 500 and 550$\upmu$m
\citep{Marshall2017} may recognize the degradation of the circalunar cycle exactly as described in our paper.
How they are affected by a decreased amplitude of the circa-monthly signal is yet to be
shown in future (chrono)biological studies.

\begin{table}
\centering
\caption[Moonlight Factor y]{Station codes and zero-frequency \NSB\ in units of $\mathsf{\magsqm}$
are shown in columns 1 and 2. Calculations of the moonlight factor (y) as defined in \cite{Bara2016} are
given in column 3}
\label{tab:moonlightfactor}
\begin{tabular}{ccc}
COD & \NSB & $y\times10^3$ \\
(1) & (2) & (3) \\
\hline \hline
    & \textit{urban}        &      \\
LGO & 17.31        & 0.10 \\
LSM & 17.51        & 0.09 \\
WEL & 17.7         & 0.10 \\
LSW & 17.73        & 0.13 \\
STW & 17.93        & 0.16 \\
TRA & 18.09        & 0.16 \\
STY & 18.27        & 0.19 \\
    & \textit{intermediate} &      \\
PAS & 18.46        & 0.21 \\
BRA & 18.54        & 0.32 \\
GRI & 18.6         & 0.23 \\
VOE & 18.62        & 0.30 \\
FRE & 18.6         & 0.36 \\
MAT & 19.05        & 0.47 \\
    & \textit{rural}        &      \\
MUN & 19.76        & 0.77 \\
GIS & 20.07        & 0.65 \\
ULI & 20.13        & 0.83 \\
KID & 20.37        & 0.85 \\
GRU & 20.39        & 1.08 \\
FEU & 20.66        & 0.87 \\
LOS & 20.86        & 0.97 \\
KRI & 20.81        & 0.78 \\
ZOE & 21.01        & 1.36 \\
BOD & 21.04        & 1.29 \\
SYN & 20.81        & 1.32 
\end{tabular}
\end{table}

%%%%%%%%%%%%%%%%%%%%%%%%%%%%%%%%%%%%%%%%%%%%%%%%%%%%%%%%%%%%%%%%%%%%%%%%%%%%%%%%%%%
%%%%%%%%%%%%%%%%%%%%%%%%%%%%%%%%%%%%%%%%%%%%%%%%%%%%%%%%%%%%%%%%%%%%%%%%%%%%%%%%%%%
\section{Summary}
We studied the circalunar periodicity via FFT analysis of night-time averages of
zenithal NSB measurements obtained during the years 2016 and 2017
through a network of 23 SQMs located in Upper Austria at a latitude of $\sim$48$^\degree$.
Models of the sky \citep{Noll2012,Jones2013} and the Moon \citep{Seidelmann1992}
were used as a reference of ideal conditions and to develop an optimal methodology
for the recognition of the circalunar periodicity. The following conclusions are drawn from our study:

\begin{itemize}
    \item A tight linear relation between \NSB\ given in $\mathsf{\magsqm}$ and the circalunar amplitude is found
    (see Figure \ref{fig:circalunarrelation}). This relation indicates that for sites with \NSB\
    brighter than 16.5 $\mathsf{\magsqm}$ the lunar rhythm practically vanishes.
    This finding implies that the circalunar rhythm is still detectable (within the broad bandpass of the SQM)
    at most places around the globe.
    
    \item However, the circalunar contrast in zenith is largely reduced compared to ALAN-free sites.
    In the state capital of Linz ($\sim$200,000 inhabitants) the Moon's contribution to zenithal \NSB\
    is reduced by a factor of $\mathsf{\sfrac{1}{9}}$. For small towns (e.g. Freistadt or Mattighofen)
    with less than 10,000 inhabitants, we find that the circalunar contrast in zenith is lowered
    by a factor of $\mathsf{\sfrac{1}{3}}$ due to ALAN and even at one of our rural sites, M\"unzkirchen,
    a village with less than 2,600 inhabitants, the circalunar zenithal contrast is reduced
    to a level of $\mathsf{\sfrac{1}{2}}$ compared to ALAN-free conditions.
    
    \item Only two of our sites, both situated in national parks (Bodinggraben and Z\"oblboden),
    show natural circalunar amplitudes.

    \item At our urban sites we further detect a strong
    seasonal signal that is linked to the amplification of anthropogenic skyglow
    during the winter months due to combined effects of enhanced albedo (due to snow)
    and a lower vegetation state (Puschnig et al. in prep).
    
    \item At urban sites we further detect a significant ($>$3 sigma) signal with a
    periodic time of $\sim$100 days. The cause of this roughly quaterly variation is
    unclear and was not reported previously.
\end{itemize}

%%%%%%%%%%%%%%%%%%%%%%%%%%%%%%%%%%%%%%%%%%%%%%%%%%%%%%%%%%%%%%%%%%%%%%%%%%%%%%%%%%%
%%%%%%%%%%%%%%%%%%%%%%%%%%%%%%%%%%%%%%%%%%%%%%%%%%%%%%%%%%%%%%%%%%%%%%%%%%%%%%%%%%%
\section*{Acknowledgements}
We dedicate this work to Thomas Posch, our most valued colleague,
mentor and dearest friend,
who passed away during the development phase of this manuscript.
This paper would have never been written without Thomas, as it was him who first introduced
Johannes and Stefan to this stunning field of research many years ago.
He will be missed by us, not only because of his
expertise and keen mind, but also because of his cordiality.

We acknowledge support by the provincial government of Upper Austria, in particular we thank
Heribert Kaineder and Martin Waslmaier for their dedication and tireless efforts to
establish a network of SQMs in Upper Austria, and share the data with the public.

We are grateful to the referee for his/her constructive input.

This research further made use of SciPy \citep{jones2001scipy} and NumPy \citep{van2011numpy},
two \texttt{Python} packages that make life as a scientist easier.

%%%%%%%%%%%%%%%%%%%%%%%%%%%%%%%%%%%%%%%%%%%%%%%%%%



\bibliographystyle{unsrtnat}
\bibliography{fft} % if your bibtex file is called example.bib

%%%%%%%%%%%%%%%%%%%%%%%%%%%%%%%%%%%%%%%%%%%%%%%%%%

%%%%%%%%%%%%%%%%% APPENDICES %%%%%%%%%%%%%%%%%%%%%
\appendix
\onecolumn

\begin{figure}
\centering
        \includegraphics[width=0.85\textwidth]{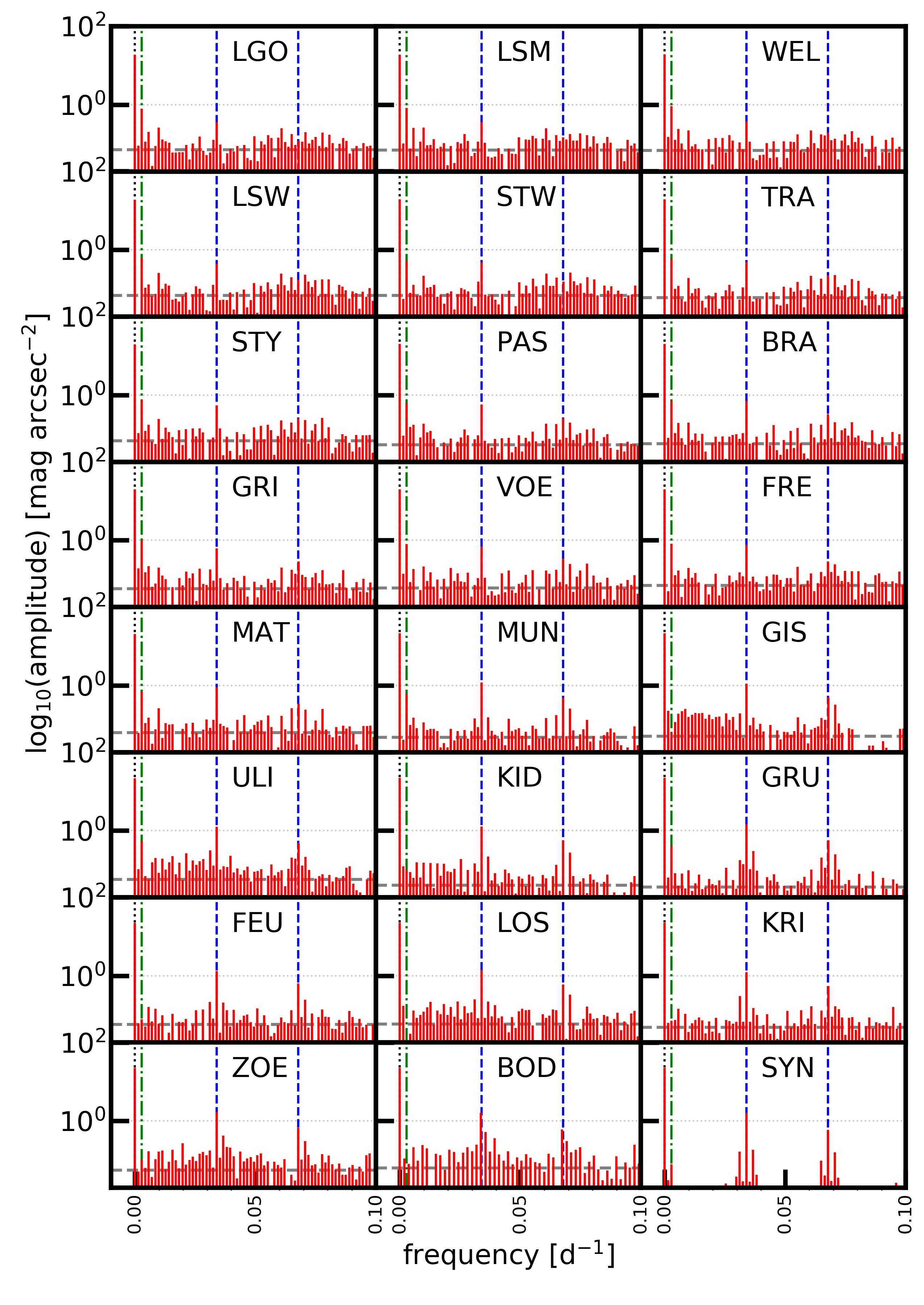}
        \caption[FFT amplitude spectra]{The panels show cutouts of the FFT amplitude spectra
        ranging from $\mathsf{0}$ to $\mathsf{0.1\,d^{-1}}$
        for 23 locations in Upper Austria, sorted by increasing
        \NSB\ values, i.e. decreasing light pollution. The amplitude spectra were obtained from a two
        year long dataset (2016-2017) of nightly \NSB\ values.
        For comparison, an amplitude spectrum of a cloud-free synthetic sky model as described in section \ref{sec:skymodel} is shown in the
        bottom right panel (SYN). The peak at zero frequency, i.e. \NSB, is indicated by a black dotted line,
        the seasonal variation $\mathsf{\frac{1}{365}\ d^{-1}}$ is marked with a green
        dash-dotted line, the expected lunar synodic frequency of $\mathsf{\frac{1}{29.5}\ d^{-1}}$ and its first harmonic
        are marked with blue, dashed lines and the mean noise level is shown as horizontal dashed line.}
        \label{fig:fft}
\end{figure}

%%%%%%%%%%%%%%%%%%%%%%%%%%%%%%%%%%%%%%%%%%%%%%%%%%%%%%%%%%%%%%%%%%%%%%%%%%%%%%%%%%%
%%%%%%%%%%%%%%%%%%%%%%%%%%%%%%%%%%%%%%%%%%%%%%%%%%%%%%%%%%%%%%%%%%%%%%%%%%%%%%%%%%%
\begin{table*}
\tiny
\centering
\caption[Summary of significant peaks in the FFT amplitude spectra]{Summary of significant ($\mathsf{\frac{S}{N}>3}$) peaks
in the FFT amplitude spectra. No windowing function was applied, instead we truncated the input time series at
the edges in order to avoid discontinuities with respect to the lunar synodic month. That way, the circalunar amplitude
could be recovered with highest accuracy. However, other amplitudes such as the one attributed to the
seasonal variation are only lower limits in most cases. We found that the application of a Hanning window leads
to an increase of the seasonal amplitude between $\mathsf{0.05}$ and $\mathsf{0.1\,\magsqm}$.}
\label{tab:fftpeaks}
\begin{tabular}{lllllllllllll}
$\mathsf{COD}$ & $\mathsf{\upnu}$ & $\mathsf{T}$ & $\mathsf{A}$ & $\mathsf{S/N}$ & $\mathsf{note}$ &  &
$\mathsf{COD}$ & $\mathsf{\upnu}$ & $\mathsf{T}$ & $\mathsf{A}$ & $\mathsf{S/N}$ & $\mathsf{note}$ \\
(1) & (2) & (3) & (4) & (5) & (6) &  & (1) & (2) & (3) & (4) & (5) & (6) \\
\hline \hline
LGO & 0.00e+00 & inf & 17.31 & 244.0 & mean &  & BRA & 0.00e+00 & inf & 18.54 & 326.3 & mean \\
LGO & 2.83e-03 & 353.0 & 0.73 & 10.3 & seasonal &  & BRA & 2.82e-03 & 354.0 & 0.67 & 11.8 & seasonal \\
LGO & 9.92e-03 & 100.9 & 0.24 & 3.4 & unknown &  & BRA & 5.65e-03 & 177.0 & 0.18 & 3.2 & seasonal (1st harm.) \\
LGO & 3.40e-02 & 29.4 & 0.34 & 4.7 & circalunar &  & BRA & 9.89e-03 & 101.1 & 0.19 & 3.3 & unknown \\
LGO & 6.09e-02 & 16.4 & 0.24 & 3.3 &  &  & BRA & 3.39e-02 & 29.5 & 0.66 & 11.7 & circalunar \\
 &  &  &  &  &  &  & BRA & 6.07e-02 & 16.5 & 0.17 & 3.1 &  \\
LSM & 0.00e+00 & inf & 17.51 & 250.6 & mean &  & BRA & 6.78e-02 & 14.8 & 0.30 & 5.3 & circalunar (1st harm.) \\
LSM & 2.82e-03 & 354.0 & 0.77 & 11.1 & seasonal &  & BRA & 7.06e-02 & 14.2 & 0.19 & 3.3 &  \\
LSM & 5.65e-03 & 177.0 & 0.24 & 3.4 & seasonal (1st harm.) &  & BRA & 7.77e-02 & 12.9 & 0.19 & 3.3 &  \\
LSM & 9.89e-03 & 101.1 & 0.24 & 3.5 & unknown &  & BRA & 1.03e-01 & 9.7 & 0.17 & 3.0 & circalunar (2nd harm.) \\
LSM & 3.39e-02 & 29.5 & 0.33 & 4.8 & circalunar &  &  &  &  &  &  &  \\
LSM & 6.07e-02 & 16.5 & 0.24 & 3.4 &  &  & FRE & 0.00e+00 & inf & 18.60 & 267.6 & mean \\
 &  &  &  &  &  &  & FRE & 2.82e-03 & 354.0 & 0.76 & 10.9 & seasonal \\
WEL & 0.00e+00 & inf & 17.70 & 263.2 & mean &  & FRE & 3.39e-02 & 29.5 & 0.71 & 10.2 & circalunar \\
WEL & 2.82e-03 & 354.0 & 0.86 & 12.8 & seasonal &  & FRE & 6.78e-02 & 14.8 & 0.27 & 3.9 & circalunar (1st harm.) \\
WEL & 5.65e-03 & 177.0 & 0.22 & 3.3 & seasonal (1st harm.) &  & FRE & 7.06e-02 & 14.2 & 0.23 & 3.3 &  \\
WEL & 9.89e-03 & 101.1 & 0.21 & 3.1 & unknown &  &  &  &  &  &  &  \\
WEL & 3.39e-02 & 29.5 & 0.36 & 5.4 & circalunar &  & GRI & 0.00e+00 & inf & 18.60 & 324.6 & mean \\
WEL & 6.07e-02 & 16.5 & 0.21 & 3.1 &  &  & GRI & 1.42e-03 & 706.0 & 0.18 & 3.1 &  \\
 &  &  &  &  &  &  & GRI & 2.83e-03 & 353.0 & 0.92 & 16.1 & seasonal \\
LSW & 0.00e+00 & inf & 17.73 & 256.2 & mean &  & GRI & 5.67e-03 & 176.5 & 0.20 & 3.5 & seasonal (1st harm.) \\
LSW & 2.82e-03 & 354.0 & 0.58 & 8.4 & seasonal &  & GRI & 9.92e-03 & 100.9 & 0.19 & 3.2 & unknown \\
LSW & 9.89e-03 & 101.1 & 0.24 & 3.5 & unknown &  & GRI & 2.69e-02 & 37.2 & 0.17 & 3.0 &  \\
LSW & 3.39e-02 & 29.5 & 0.41 & 6.0 & circalunar &  & GRI & 3.40e-02 & 29.4 & 0.57 & 10.0 & circalunar \\
LSW & 6.07e-02 & 16.5 & 0.23 & 3.3 &  &  & GRI & 6.09e-02 & 16.4 & 0.18 & 3.2 &  \\
LSW & 7.06e-02 & 14.2 & 0.21 & 3.1 &  &  & GRI & 6.80e-02 & 14.7 & 0.27 & 4.7 & circalunar (1st harm.) \\
 &  &  &  &  &  &  & GRI & 1.03e-01 & 9.7 & 0.17 & 3.0 & circalunar (2nd harm.) \\
STW & 0.00e+00 & inf & 17.93 & 259.4 & mean &  &  &  &  &  &  &  \\
STW & 2.83e-03 & 353.0 & 0.49 & 7.0 & seasonal &  & VOE & 0.00e+00 & inf & 18.62 & 309.0 & mean \\
STW & 3.40e-02 & 29.4 & 0.45 & 6.5 & circalunar &  & VOE & 2.82e-03 & 354.0 & 0.74 & 12.2 & seasonal \\
STW & 6.09e-02 & 16.4 & 0.23 & 3.3 &  &  & VOE & 9.89e-03 & 101.1 & 0.20 & 3.3 & unknown \\
STW & 7.08e-02 & 14.1 & 0.24 & 3.5 &  &  & VOE & 2.12e-02 & 47.2 & 0.18 & 3.0 &  \\
 &  &  &  &  &  &  & VOE & 3.39e-02 & 29.5 & 0.64 & 10.6 & circalunar \\
TRA & 0.00e+00 & inf & 18.09 & 296.7 & mean &  & VOE & 6.78e-02 & 14.8 & 0.32 & 5.3 & circalunar (1st harm.) \\
TRA & 2.82e-03 & 354.0 & 0.55 & 9.1 & seasonal &  & VOE & 7.06e-02 & 14.2 & 0.23 & 3.8 &  \\
TRA & 3.39e-02 & 29.5 & 0.46 & 7.5 & circalunar &  & VOE & 7.77e-02 & 12.9 & 0.24 & 3.9 &  \\
TRA & 6.07e-02 & 16.5 & 0.20 & 3.3 &  &  & VOE & 1.03e-01 & 9.7 & 0.18 & 3.0 & circalunar (2nd harm.) \\
TRA & 6.78e-02 & 14.8 & 0.21 & 3.4 & circalunar (1st harm.) &  &  &  &  &  &  &  \\
TRA & 7.06e-02 & 14.2 & 0.21 & 3.5 &  &  & MAT & 0.00e+00 & inf & 19.05 & 305.3 & mean \\
 &  &  &  &  &  &  & MAT & 2.83e-03 & 353.0 & 0.65 & 10.4 & seasonal \\
STY & 0.00e+00 & inf & 18.27 & 268.6 & mean &  & MAT & 9.92e-03 & 100.9 & 0.25 & 3.9 & unknown \\
STY & 2.82e-03 & 354.0 & 0.69 & 10.1 & seasonal &  & MAT & 3.40e-02 & 29.4 & 0.83 & 13.3 & circalunar \\
STY & 9.89e-03 & 101.1 & 0.23 & 3.3 & unknown &  & MAT & 6.52e-02 & 15.3 & 0.24 & 3.9 &  \\
STY & 3.39e-02 & 29.5 & 0.50 & 7.4 & circalunar &  & MAT & 6.80e-02 & 14.7 & 0.32 & 5.1 & circalunar (1st harm.) \\
STY & 6.07e-02 & 16.5 & 0.21 & 3.1 &  &  & MAT & 7.08e-02 & 14.1 & 0.22 & 3.6 &  \\
STY & 6.78e-02 & 14.8 & 0.23 & 3.5 & circalunar (1st harm.) &  & MAT & 7.79e-02 & 12.8 & 0.23 & 3.8 &  \\
STY & 7.06e-02 & 14.2 & 0.21 & 3.1 &  &  & MAT & 1.78e-01 & 5.6 & 0.19 & 3.1 &  \\
STY & 7.77e-02 & 12.9 & 0.24 & 3.6 &  &  &  &  &  &  &  &  \\
 &  &  &  &  &  &  & MUN & 0.00e+00 & inf & 19.76 & 421.4 & mean \\
PAS & 0.00e+00 & inf & 18.46 & 344.1 & mean &  & MUN & 2.82e-03 & 354.0 & 0.56 & 12.0 & seasonal \\
PAS & 2.82e-03 & 354.0 & 0.57 & 10.7 & seasonal &  & MUN & 3.39e-02 & 29.5 & 1.10 & 23.4 & circalunar \\
PAS & 5.65e-03 & 177.0 & 0.16 & 3.0 & seasonal (1st harm.) &  & MUN & 3.67e-02 & 27.2 & 0.14 & 3.0 &  \\
PAS & 9.89e-03 & 101.1 & 0.17 & 3.2 & unknown &  & MUN & 6.50e-02 & 15.4 & 0.16 & 3.5 &  \\
PAS & 3.39e-02 & 29.5 & 0.53 & 9.9 & circalunar &  & MUN & 6.78e-02 & 14.8 & 0.45 & 9.6 & circalunar (1st harm.) \\
PAS & 6.07e-02 & 16.5 & 0.17 & 3.2 &  &  & MUN & 7.06e-02 & 14.2 & 0.24 & 5.1 &  \\
PAS & 6.50e-02 & 15.4 & 0.17 & 3.1 &  &  & MUN & 1.02e-01 & 9.8 & 0.19 & 4.1 & circalunar (2nd harm.) \\
PAS & 6.78e-02 & 14.8 & 0.24 & 4.5 & circalunar (1st harm.) &  &  &  &  &  &  &  \\
PAS & 7.06e-02 & 14.2 & 0.18 & 3.4 &  &  &  &  &  &  &  &  \\
\end{tabular}
\end{table*}

\begin{table*}
\tiny
\begin{tabular}{lllllllllllll}
$\mathsf{COD}$ & $\mathsf{\upnu}$ & $\mathsf{T}$ & $\mathsf{A}$ & $\mathsf{S/N}$ & $\mathsf{note}$ &  &
$\mathsf{COD}$ & $\mathsf{\upnu}$ & $\mathsf{T}$ & $\mathsf{A}$ & $\mathsf{S/N}$ & $\mathsf{note}$ \\
(1) & (2) & (3) & (4) & (5) & (6) &  & (1) & (2) & (3) & (4) & (5) & (6) \\
\hline \hline
GIS & 0.00e+00 & inf & 20.07 & 400.5 & mean &  & FEU & 0.00e+00 & inf & 20.66 & 362.6 & mean \\
GIS & 1.42e-03 & 706.0 & 0.21 & 4.2 &  &  & FEU & 3.11e-02 & 32.2 & 0.20 & 3.5 &  \\
GIS & 5.67e-03 & 176.5 & 0.18 & 3.6 & seasonal (1st harm.) &  & FEU & 3.39e-02 & 29.5 & 1.22 & 21.5 & circalunar \\
GIS & 7.08e-03 & 141.2 & 0.21 & 4.1 &  &  & FEU & 3.67e-02 & 27.2 & 0.19 & 3.4 &  \\
GIS & 8.50e-03 & 117.7 & 0.23 & 4.7 &  &  & FEU & 6.78e-02 & 14.8 & 0.60 & 10.5 & circalunar (1st harm.) \\
GIS & 1.13e-02 & 88.2 & 0.16 & 3.3 &  &  & FEU & 7.06e-02 & 14.2 & 0.23 & 4.0 &  \\
GIS & 1.27e-02 & 78.4 & 0.18 & 3.7 &  &  & FEU & 1.02e-01 & 9.8 & 0.24 & 4.2 & circalunar (2nd harm.) \\
GIS & 1.42e-02 & 70.6 & 0.18 & 3.6 &  &  & FEU & 1.24e-01 & 8.0 & 0.19 & 3.3 &  \\
GIS & 1.56e-02 & 64.2 & 0.18 & 3.7 &  &  &  &  &  &  &  &  \\
GIS & 1.84e-02 & 54.3 & 0.17 & 3.3 &  &  & KRI & 0.00e+00 & inf & 20.81 & 428.6 & mean \\
GIS & 2.55e-02 & 39.2 & 0.18 & 3.6 &  &  & KRI & 3.12e-02 & 32.1 & 0.28 & 5.8 &  \\
GIS & 3.12e-02 & 32.1 & 0.18 & 3.6 &  &  & KRI & 3.40e-02 & 29.4 & 1.16 & 23.9 & circalunar \\
GIS & 3.40e-02 & 29.4 & 1.02 & 20.3 & circalunar &  & KRI & 6.09e-02 & 16.4 & 0.16 & 3.2 &  \\
GIS & 6.80e-02 & 14.7 & 0.47 & 9.4 & circalunar (1st harm.) &  & KRI & 6.80e-02 & 14.7 & 0.51 & 10.6 & circalunar (1st harm.) \\
GIS & 7.08e-02 & 14.1 & 0.30 & 6.0 &  &  & KRI & 7.08e-02 & 14.1 & 0.16 & 3.2 &  \\
GIS & 1.02e-01 & 9.8 & 0.19 & 3.7 & circalunar (2nd harm.) &  & KRI & 9.49e-02 & 10.5 & 0.15 & 3.1 &  \\
 &  &  &  &  &  &  & KRI & 1.02e-01 & 9.8 & 0.19 & 3.9 & circalunar (2nd harm.) \\
ULI & 0.00e+00 & inf & 20.13 & 353.8 & mean &  & KRI & 1.29e-01 & 7.8 & 0.16 & 3.3 &  \\
ULI & 2.83e-03 & 353.0 & 0.50 & 8.8 & seasonal &  &  &  &  &  &  &  \\
ULI & 8.50e-03 & 117.7 & 0.19 & 3.3 &  &  & LOS & 0.00e+00 & inf & 20.86 & 355.1 & mean \\
ULI & 1.13e-02 & 88.2 & 0.18 & 3.1 &  &  & LOS & 1.27e-02 & 78.7 & 0.20 & 3.4 &  \\
ULI & 1.56e-02 & 64.2 & 0.21 & 3.7 &  &  & LOS & 2.40e-02 & 41.6 & 0.21 & 3.5 &  \\
ULI & 2.12e-02 & 47.1 & 0.25 & 4.3 &  &  & LOS & 3.11e-02 & 32.2 & 0.24 & 4.0 &  \\
ULI & 2.83e-02 & 35.3 & 0.18 & 3.1 &  &  & LOS & 3.39e-02 & 29.5 & 1.30 & 22.0 & circalunar \\
ULI & 3.12e-02 & 32.1 & 0.29 & 5.2 &  &  & LOS & 3.67e-02 & 27.2 & 0.20 & 3.5 &  \\
ULI & 3.40e-02 & 29.4 & 1.16 & 20.4 & circalunar &  & LOS & 6.78e-02 & 14.8 & 0.56 & 9.5 & circalunar (1st harm.) \\
ULI & 3.97e-02 & 25.2 & 0.21 & 3.8 &  &  & LOS & 7.06e-02 & 14.2 & 0.31 & 5.2 &  \\
ULI & 6.52e-02 & 15.3 & 0.19 & 3.4 &  &  &  &  &  &  &  &  \\
ULI & 6.66e-02 & 15.0 & 0.18 & 3.1 &  &  & ZOE & 0.00e+00 & inf & 21.01 & 378.2 & mean \\
ULI & 6.80e-02 & 14.7 & 0.45 & 7.9 & circalunar (1st harm.) &  & ZOE & 1.55e-02 & 64.4 & 0.17 & 3.1 &  \\
ULI & 7.08e-02 & 14.1 & 0.20 & 3.5 &  &  & ZOE & 1.98e-02 & 50.6 & 0.25 & 4.5 &  \\
 &  &  &  &  &  &  & ZOE & 3.39e-02 & 29.5 & 1.55 & 28.0 & circalunar \\
KID & 0.00e+00 & inf & 20.37 & 506.3 & mean &  & ZOE & 3.67e-02 & 27.2 & 0.39 & 7.1 &  \\
KID & 7.06e-03 & 141.6 & 0.14 & 3.6 &  &  & ZOE & 3.81e-02 & 26.2 & 0.20 & 3.7 &  \\
KID & 9.89e-03 & 101.1 & 0.14 & 3.5 & unknown &  & ZOE & 3.95e-02 & 25.3 & 0.19 & 3.5 &  \\
KID & 1.27e-02 & 78.7 & 0.14 & 3.5 &  &  & ZOE & 6.78e-02 & 14.8 & 0.62 & 11.2 & circalunar (1st harm.) \\
KID & 1.55e-02 & 64.4 & 0.14 & 3.4 &  &  & ZOE & 7.06e-02 & 14.2 & 0.28 & 5.1 &  \\
KID & 1.84e-02 & 54.5 & 0.13 & 3.3 &  &  &  &  &  &  &  &  \\
KID & 2.54e-02 & 39.3 & 0.17 & 4.3 &  &  & BOD & 0.00e+00 & inf & 21.04 & 329.6 & mean \\
KID & 3.11e-02 & 32.2 & 0.14 & 3.4 &  &  & BOD & 5.62e-03 & 178.0 & 0.20 & 3.1 &  \\
KID & 3.39e-02 & 29.5 & 1.19 & 29.6 & circalunar &  & BOD & 9.36e-03 & 106.8 & 0.23 & 3.5 &  \\
KID & 3.67e-02 & 27.2 & 0.20 & 5.0 &  &  & BOD & 2.81e-02 & 35.6 & 0.20 & 3.1 &  \\
KID & 6.50e-02 & 15.4 & 0.12 & 3.1 &  &  & BOD & 3.18e-02 & 31.4 & 0.23 & 3.6 &  \\
KID & 6.78e-02 & 14.8 & 0.52 & 13.0 & circalunar (1st harm.) &  & BOD & 3.37e-02 & 29.7 & 1.51 & 23.6 & circalunar \\
KID & 7.06e-02 & 14.2 & 0.25 & 6.3 &  &  & BOD & 3.56e-02 & 28.1 & 0.48 & 7.6 &  \\
KID & 1.02e-01 & 9.8 & 0.20 & 4.9 & circalunar (2nd harm.) &  & BOD & 3.93e-02 & 25.4 & 0.34 & 5.3 &  \\
KID & 1.34e-01 & 7.5 & 0.12 & 3.0 & circalunar (3rd harm.) &  & BOD & 6.74e-02 & 14.8 & 0.56 & 8.7 & circalunar (1st harm.) \\
 &  &  &  &  &  &  & BOD & 6.93e-02 & 14.4 & 0.29 & 4.5 &  \\
GRU & 0.00e+00 & inf & 20.39 & 568.6 & mean &  & BOD & 7.49e-02 & 13.3 & 0.20 & 3.1 &  \\
GRU & 2.83e-03 & 353.0 & 0.38 & 10.6 & seasonal &  & BOD & 9.74e-02 & 10.3 & 0.23 & 3.6 &  \\
GRU & 3.12e-02 & 32.1 & 0.17 & 4.6 &  &  & BOD & 1.01e-01 & 9.9 & 0.20 & 3.1 & circalunar (2nd harm.) \\
GRU & 3.26e-02 & 30.7 & 0.13 & 3.6 &  &  &  &  &  &  &  &  \\
GRU & 3.40e-02 & 29.4 & 1.34 & 37.3 & circalunar &  &  &  &  &  &  &  \\
GRU & 3.68e-02 & 27.2 & 0.28 & 7.7 &  &  &  &  &  &  &  &  \\
GRU & 6.52e-02 & 15.3 & 0.19 & 5.3 &  &  &  &  &  &  &  &  \\
GRU & 6.80e-02 & 14.7 & 0.53 & 14.8 & circalunar (1st harm.) &  &  &  &  &  &  &  \\
GRU & 7.08e-02 & 14.1 & 0.23 & 6.4 &  &  &  &  &  &  &  &  \\
GRU & 1.02e-01 & 9.8 & 0.15 & 4.1 & circalunar (2nd harm.) &  &  &  &  &  &  &  \\
GRU & 1.05e-01 & 9.5 & 0.11 & 3.1 &  &  &  &  &  &  &  &  \\
\end{tabular}
\end{table*}

%%%%%%%%%%%%%%%%%%%%%%%%%%%%%%%%%%%%%%%%%%%%%%%%%%%%%%%%%%%%%%%%%%%%%%%%%%%%%%%%%%%
%%%%%%%%%%%%%%%%%%%%%%%%%%%%%%%%%%%%%%%%%%%%%%%%%%%%%%%%%%%%%%%%%%%%%%%%%%%%%%%%%%%
\begin{figure}
\centering
        \includegraphics[width=0.85\textwidth]{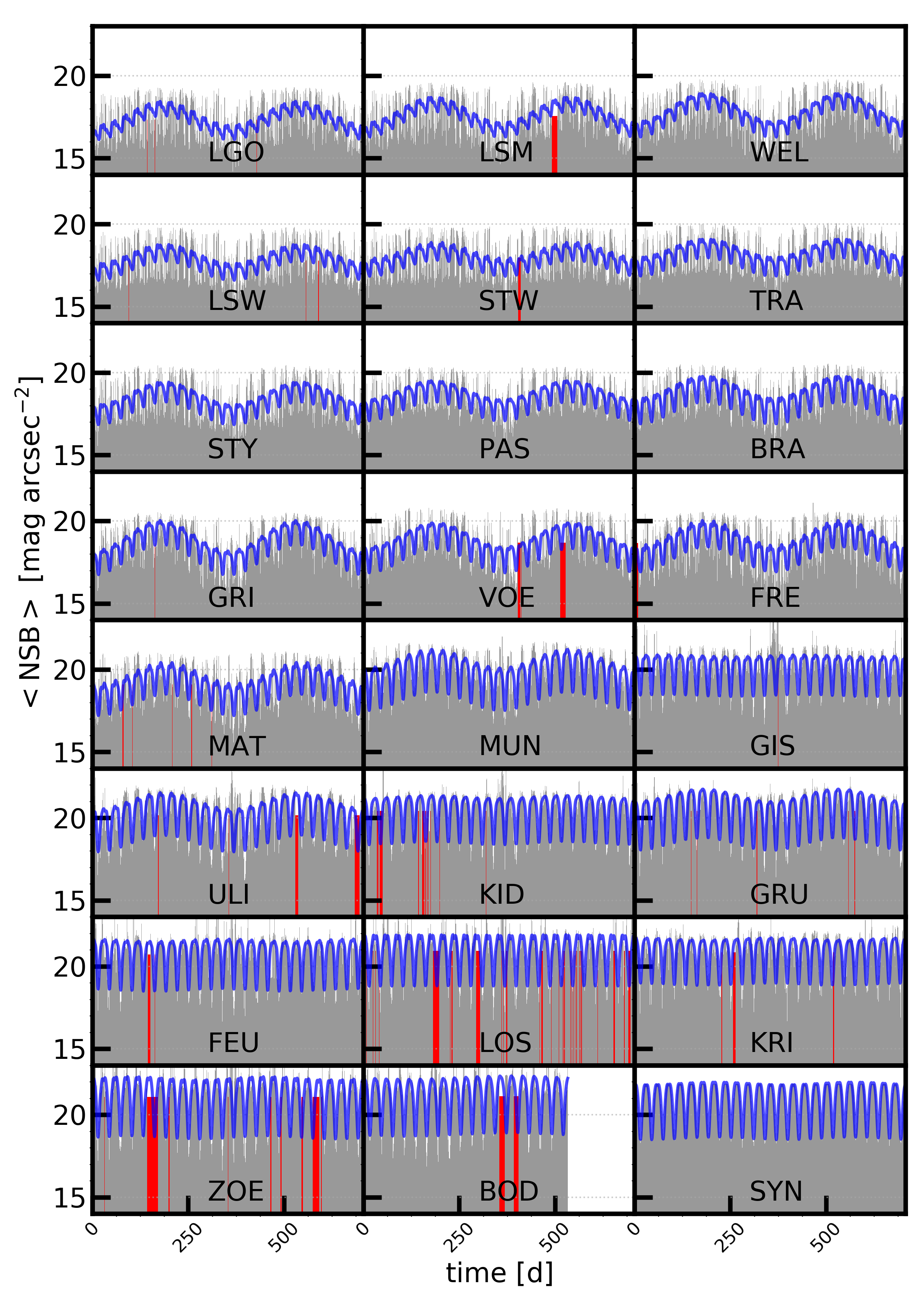}
        \caption[Time series of SQM data and iFFT of identified peaks]{The
        \textit{gray shaded areas} in the panels show \NSB\ time series for 2016 and 2017
        obtained from 23 SQM stations in Upper Austria and from a
        cloud-free synthetic sky model (SYN) as described in section \ref{sec:skymodel}.
        On top of that,
        the result of an inverse FFT (iFFT) of identified frequency components is shown as
        \textit{blue line}. The iFFT frequencies correspond to
        the mean brightness level, the circalunar rhythm plus its first two harmonics as well as  
        a yearly cycle, i.e. bright winters and dark summers. This cycle is driven by an increase of
        overcast skies during the winter months and an amplification of light pollution
        by clouds. Data gaps
        that were set to the mean \NSB\ value are marked in \textit{red}.}
        \label{fig:ifft}
\end{figure}
\begin{figure}
\centering
        \includegraphics[width=0.85\textwidth]{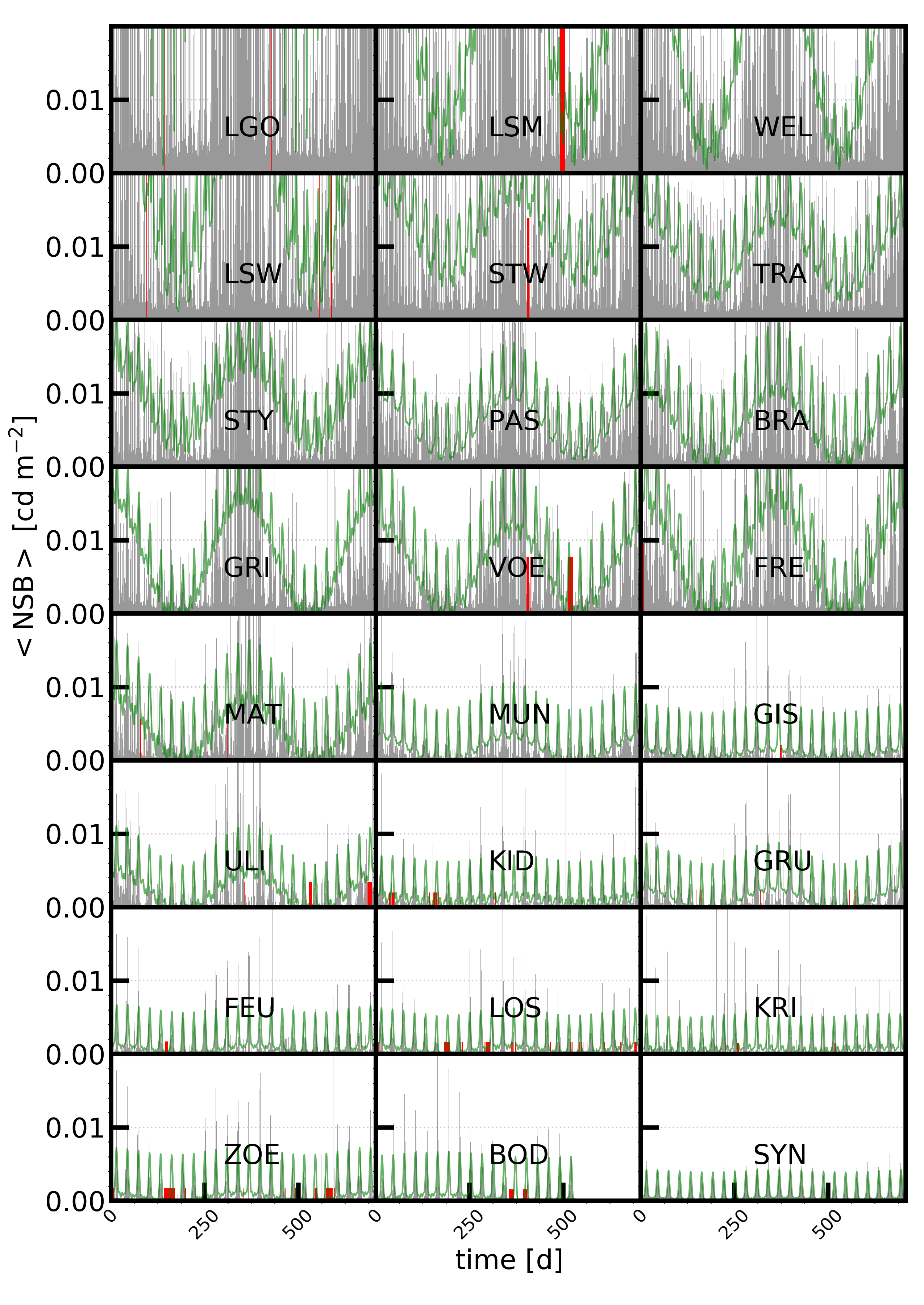}
        \caption[Time series of SQM data and iFFT of identified peaks]{The
        \textit{gray shaded areas} in the panels show \NSB\ time series for 2016 and 2017
        obtained from 23 SQM stations in Upper Austria and from a
        cloud-free synthetic sky model (SYN) as described in section \ref{sec:skymodel}.
        On top of that,
        the result of an inverse FFT (iFFT) of identified frequency components is shown as
        \textit{blue line}. The iFFT frequencies correspond to
        the mean brightness level, the circalunar rhythm plus its first two harmonics as well as  
        a yearly cycle, i.e. bright winters and dark summers. This cycle is driven by an increase of
        overcast skies during the winter months and an amplification of light pollution
        by clouds. Data gaps
        that were set to the mean \NSB\ value are marked in \textit{red}.}
        \label{fig:ifft_cdm}
\end{figure}
%%%%%%%%%%%%%%%%%%%%%%%%%%%%%%%%%%%%%%%%%%%%%%%%%%

% Don't change these lines
\end{document}